\documentclass{aa}

\usepackage{graphicx}
\usepackage{txfonts}

\usepackage[colorlinks=true,urlcolor=blue,citecolor=red,linkcolor=red,bookmarks=true]{hyperref}

\usepackage{amsmath}
\usepackage{gensymb}
\usepackage{makeidx}
\def\e#1{{\em #1}}

\begin{document}

\title{The young Adelaide family: Possible sibling to Datura?}

\author{D. Vokrouhlick\'y\inst{1},
        B. Novakovi\'{c}\inst{2},
        and D. Nesvorn\'y\inst{3}}

\titlerunning{The young Adelaide family: A sibling to Datura?}
\authorrunning{Vokrouhlick\'y et~al.}

\institute{Institute of Astronomy, Charles University, V~Hole\v{s}ovi\v{c}k\'ach 2,
           CZ-180~00 Prague 8, Czech Republic \\ \email{vokrouhl@cesnet.cz}
      \and
           Department of Astronomy, Faculty of Mathematics, University of Belgrade,
           Studentski trg 16, 11000 Belgrade, Serbia 
      \and
           Southwest Research Institute, 1050 Walnut St, Suite 300,
           Boulder, CO 80302, USA}

\date{Received: \today ; accepted: ???}

\abstract
{Very young asteroid families may record processes that accompanied their formation
 in the most pristine way. This makes analysis of this special class particularly
 interesting.}
{We  studied the very young Adelaide family in the inner part of the main belt. This
 cluster is extremely close to the previously known Datura family in the space of
 proper orbital elements and their ages overlap. As a result, we investigated the
 possibility of a causal relationship between the two families.}
{We identified Adelaide family members in the up-to-date catalogue of asteroids. By
 computing their proper orbital elements we inferred the family structure. Backward 
 orbital integration of selected members allowed us to determine the age of the
 family.}
{The largest fragment (525)~Adelaide, an S-type asteroid about $10$~km in size, is
 accompanied by 50 sub-kilometre  fragments. This family is a typical example of a
 cratering event. The very tiny extent in the semi-major axis minimises chances
 that some significant mean motion resonances influence the dynamics of its
 members, though we recognise that part
 of the Adelaide family is affected by weak, three-body resonances. Weak chaos
 is also produced by distant encounters with Mars. Simultaneous convergence of
 longitude of node for the orbits of six selected members to that of (525)~Adelaide
 constrains the Adelaide family age to $536\pm 12$~kyr (formal solution). While
 suspiciously overlapping with the age of the Datura family, we find it unlikely
 that the formation events of the two families are causally linked. In all likelihood,
 the similarity of their ages is just a coincidence.}
{}


\maketitle

%------------------------------------------------------------------------------------

% SEC 1 %%%%%%%%%%%%%%%%%%%%%%%%%%%%%%%%%%%%%%%%%%%%%%%%%%%%%%%%%%%%%%%%%%%%%%%%%%%%%
\section{Introduction}
About 50 asteroid families, for the most part products of the collisions of two
parent asteroids, have been identified in the main belt to date
\citep[e.g.][]{netal2015,metal2015}. They
represent a wide variety of all kinds of sizes, shapes, spectral types, and/or ages. Large
and old families are impressive as they represent the outcomes of enormously energetic
collisions. An analysis of their fragments may give us information about the interiors of giant
planetesimals in terms of chemical composition and mechanical properties governing
the fragmentation process. They are also crucial tracers of the early epochs when the 
terrestrial planets experienced increased impact flux. However, the analysis of
large families may hide difficulties. This is because over the eons since their formation,
many dynamical and physical processes might have changed their properties. In addition, 
large portions of the families may have been lost through major resonances.
Disentangling the histories of large families may
 thus be a great challenge and subject to large uncertainties.

In contrast, the analysis of young families may be more straightforward   because the
dynamical and physical processes that  make them evolve have  not had enough time
to operate. Therefore, young families are potentially left in a much more pristine state,
providing better clues to their formation. Additionally, families younger than
about $10$~Myr offer a unique possibility of dating their origins using the direct integration
of heliocentric orbits of their members backwards in time. This approach, first applied
to the Karin family by \citet{karin2002}, may result in age estimations accurate to about
$\leq 10$\% in relative terms, impossible to achieve for older families. This information
may in turn be crucial for potentially linking the family's formation events to the
accretion record of terrestrial planets \citep[e.g.][]{fetal2006}.  A possible downside
in the study of young young families consists of the
typically smaller size of their parent bodies \citep[which are statistically more likely
to collisionally disrupt; e.g.][]{betal2005}. As a result, fragments forming such families
are also typically small objects, many of which may not yet have been discovered by current sky
surveys. This aspect is the most severe in very young asteroid families with  ages less
than $1$~Myr, which often consist of only a few known members. The first examples in this
category were discovered little more than a decade ago \citep{datura2006,nv2006}. Only 
two rare cases of very young families with a numerous population of known fragments have 
been studied in detail so far, the Datura family \citep[e.g.][]{datura2009,datura2017} and
the Schulhof family \citep[e.g.][]{vetal2011,vetal2016}.

\citet{ade2019} reported the discovery of a very young family in the vicinity of the
asteroid (525)~Adelaide, thence the Adelaide family. In this brief note the authors used
backward orbital integration of identified members to infer an approximate age
of $\simeq 500$~kyr from the dispersion of the nodal longitudes. However, further details
about this cluster, including a list of the identified members, were not given. Our
interest in the Adelaide family was primarily driven by the suggested age,
which seems suspiciously similar to that of Datura family \citep{datura2009}.
Moreover, both families are located in a very similar orbital zone. We considered the
proper orbital elements of (1270)~Datura (semi-major axis $a_{\rm P}=2.2347$~au,
eccentricity $e_{\rm P}=0.1535$, and sine of inclination $\sin I_{\rm P}=0.0920$),
and compared them with those of (525)~Adelaide (semi-major axis $a_{\rm P}=2.2452$~au,
eccentricity $e_{\rm P}=0.1487$, and sine of inclination $\sin I_{\rm P}=0.1170$).
After we reviewed the currently known population of members in the Adelaide family
and determined its age using a somewhat more accurate approach (Sect.~\ref{fam}), we 
discuss its hypothetical relation to the Datura family (Sect.~\ref{disc}). In particular,
we considered a case of a causal relationship between them, such that, for instance,  the
Adelaide family formed first and soon afterwards a coherent stream of its
fragments hit the parent asteroid of the Datura family. If true, this would be
the first chain reaction between asteroids ever described.

% SEC 2 %%%%%%%%%%%%%%%%%%%%%%%%%%%%%%%%%%%%%%%%%%%%%%%%%%%%%%%%%%%%%%%%%%%%%%%%%%%%%
\section{Adelaide family} \label{fam}

\subsection{The largest fragment} \label{lf}
The largest fragment in this cluster, (525)~Adelaide, is the only asteroid
with some physical characterisation available. \citet{metal2014} analysed
WISE/NEOWISE infrared observations and reported Adelaide's size $D=9.33\pm 0.24$~km
and geometric albedo $p_V=0.22\pm0.05$. These results assume an absolute  magnitude
$H=12.4$~mag in the visible band. More up-to-date absolute magnitude determinations
across all standard databases (such as MPC, JPL, or AstDyS) indicate a slightly
smaller value of absolute magnitude: $H\simeq 12.1$. The difference
of about $0.3$~magnitude is a characteristic scatter in this parameter reported
by \citet{petal2012}, who also noted that near $H\simeq 12$~mag the systematic
offset of the database-reported values should be small (becoming as large as
$-0.5$~mag for smaller objects). It is therefore possible that the true
size of (525)~Adelaide is slightly larger, its geometric albedo slightly higher,
or a combination of both. The answer will be found when photometrically calibrated
data of this asteroid are taken. Luckily, this issue is not critical for our
analysis. With a safe margin we can assume Adelaide's size to be  between
$9$ and $11.5$~km, and the geometric albedo between $0.22$ and $0.29$. This albedo
range closely matches the characteristic value in the Flora family region
of the inner main belt. \citet{netal2015} identified (525)~Adelaide as
a member in the Flora family despite the proper inclination value of
its orbit being at the high end of the Flora family members. Nevertheless,
whether Adelaide is a fragment from the Flora-family formation event, which 
took part more than 1 Gyr ago \citep[e.g.][]{vetal2017}, is again not
a crucial issue for our study. The confirmation of   this albedo
value also comes   from broad-band photometry of (525)~Adelaide taken by
the Sloan Digital Sky Survey (SDSS) project. Considering the methodology in
\citet{parker2008}, we used the SDSS-based observations to infer its  colour
indexes $a^* = 0.108$ and $i-z=-0.08$. Given the information in Fig.~3 of
\citet{parker2008} we concluded that these values are characteristic of
S-complex asteroids. This is a prevalent spectral type in the Flora zone,
and the characteristic albedo values of these objects very closely match
those given for (525)~Adelaide.

From what we know we can thus assume
(525)~Adelaide is a typical S-type object in its orbital region. This
information also helps to estimate its bulk density to be between $2.2$ and
$3.2$~g~cm$^{-3}$ \citep[e.g.][]{setal2015}. As a result, a characteristic value
of the escape velocity from (525)~Adelaide would safely be in the $5$ to
$10$~m~s$^{-1}$ range of values. Finally, \cite{pilcher2014} photometrically
observed (525)~Adelaide during its 2014 apparition. From these data, he inferred the
rotation period to be $P=19.967\pm 0.001$~hr and rather low amplitude $0.35\pm 0.03$~mag
of the light curve. This indicates Adelaide is a slowly rotating asteroid 
likely having a near-spherical shape, justifying the above-mentioned simple estimate
of the escape velocity (based on a spherical shape assumption).

\subsection{Family identification} \label{iden}
Asteroid families are traditionally identified as statistically significant
clusters in the space of proper orbital elements: semi-major axis ($a_{\rm P}$),
eccentricity ($e_{\rm P}$), and sine of inclination ($\sin I_{\rm P}$)
\citep[e.g.][]{ketal2002}. The hierarchical clustering method (HCM) is the
most often adopted tool to discern these clusters (though there are also other,
less employed approaches) and to evaluate their statistical significance
\citep[e.g.][]{bz2002,netal2015}. This requires us to introduce a metric function
in the three-dimensional space of proper elements in order to evaluate the distance
between two asteroid orbits, and to analyse both the family population and the
background populations of asteroids. It was realised that a problem may occur
for very young families with only a limited number of known members; in these
cases it may be difficult to prove the
statistical significance of the family among the plethora of interloping background
objects. \citet{datura2006} and \citet{nv2006} observed that the problem
occurs due to the low dimensionality of the proper element space, and proposed
that working in a higher dimensional space may help solve the issue. At the
same time, they noted that collisionally born clusters initially have very
tightly clustered longitude of node $\Omega$ and perihelion $\varpi$, in addition to
semi-major axis $a$, eccentricity $e$, and inclination $I$, and that it takes only
about $1-2$~Myr to make these secular angles disperse into a whole range of possible
values. Therefore, very young families (aged less than $1-2$~Myr) could be identified
directly in a five-dimensional space of osculating orbital elements, excluding only
longitude in orbit \citep[which disperses much faster, on approximately a thousand-year timescale;
see also][where the concept of very young families is also discussed]{rp2018}.
In order to follow classical lines of family identification, \citet{nv2006} 
introduced an empirical metric in the $(a,e,I,\Omega,\varpi)$ space and used the
HCM method. However, the metric extension to secular angles $(\Omega,\varpi)$ is
only approximate \citep[see also][]{retal2011}, and this could introduce 
uncertainties. For this reason, we actually decided to use a more straightforward
but robust approach.
% FIG %%%%%%%%%%%%%%%%%%%%%%%%%%%%%%%%%%%%%%%%%%%%%%%%%%%%%%%%%%%%%%%%%%%%%%%%%%%%%%%%
\begin{figure}[t]
 \begin{center} 
 \includegraphics[width=0.49\textwidth]{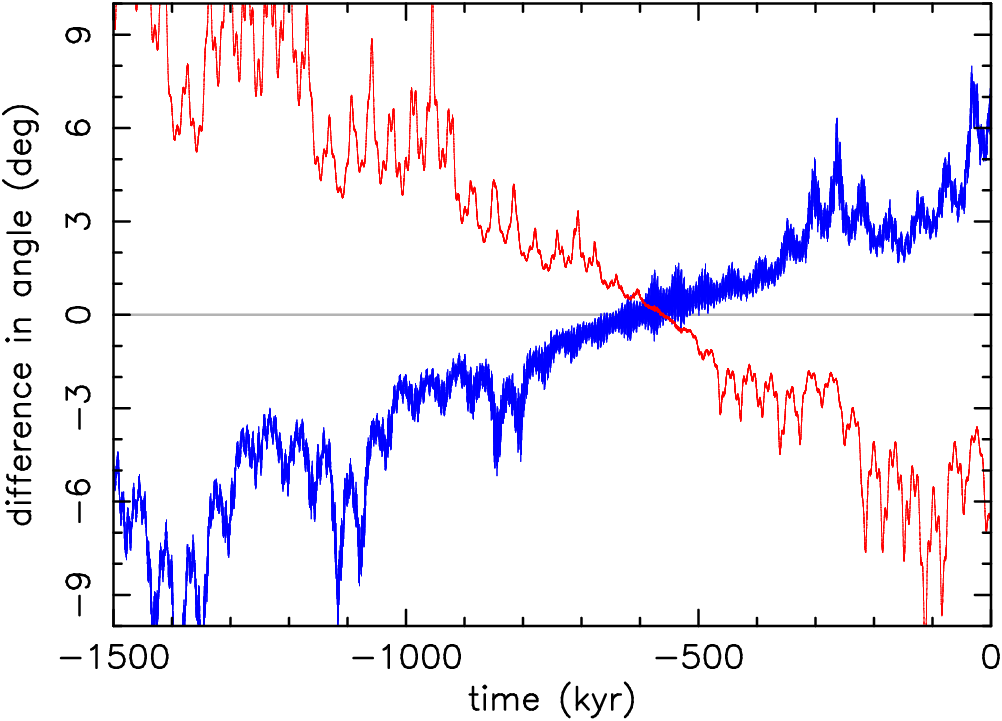}
 \end{center}
 \caption{\label{f1}
  Orbital convergence of secular angles for the nominal orbits of (452322) 2000~GG121
  and (525)~Adelaide: (i) difference in longitude of node $\delta
  \Omega=\Omega_{452322}-\Omega_{525}$ (red line), (ii) difference in longitude
  of perihelion $\delta\varpi=\varpi_{452322}-\varpi_{525}$ (blue line) (osculating values in
  both cases). The abscissa is time to the past (in kyr). At approximately $500$~kyr
  ago, both   secular angles of (452322) 2000~GG121 converge to the respective values
  of (525)~Adelaide.}
\end{figure}
%%%%%%%%%%%%%%%%%%%%%%%%%%%%%%%%%%%%%%%%%%%%%%%%%%%%%%%%%%%%%%%%%%%%%%%%%%%%%%%%%%%%%%

We constructed a simple box-zone in the space of the osculating elements $(a,e,I,
\Omega,\varpi)$ around the nominal orbit of (525)~Adelaide. The box was defined
using uncorrelated variation in each of these elements by the following
values: semi-major axis $\pm 0.01$~au, eccentricity $\pm 0.01$, inclination $\pm
0.1^\circ$, longitude of node and argument of perihelion $\pm 30^\circ$. These
values are far larger than the orbital uncertainty of (525)~Adelaide, which
justifies   using its nominal (best-fit) orbit at this stage. We used the {\tt MPCORB.DAT}
database of asteroid orbits provided by the Minor Planet Center, which contained
more than a million entries as of February~2021. We found $52$ objects, including
(525)~Adelaide, in our target box. Importantly, these asteroids are not distributed
uniformly in this zone, but are rather tightly clustered near its centre. As an example,
the longitude of node values are clustered around Adelaide's value within the
interval $-7^\circ$ to $+2^\circ$. This is much less than the box limits $\pm 30^\circ$.
Similarly, all semi-major axes values are within the interval $-0.002$~au and $+0.002$~au  
around Adelaide's value, while the target box had a $\pm 0.01$~au allowance. In both
cases the orbital parameter values occupy about one-fifth of the target zone. A similar
situation occurs for the remaining three elements. Another important aspect consists
of a correlation between various elements in the box. For instance, the longitude of node
is anticorrelated with the longitude of perihelion for the objects found. Similar
anticorrelation occurs between the longitude of node and inclination, and the longitude of
perihelion and eccentricity. These correlations indicate that this group of
objects has something in common, rather than being a random sample of asteroids
\citep[for the correlation  of the secular angles   in the case of a young Datura
family, see  Fig.~12 in][]{datura2017}. Finally, we note that asteroid
(159941) 2005~WV178 is systematically at the border of the cluster in all elements,
and therefore it is a suspected interloper.
% FIG %%%%%%%%%%%%%%%%%%%%%%%%%%%%%%%%%%%%%%%%%%%%%%%%%%%%%%%%%%%%%%%%%%%%%%%%%%%%%%%%
\begin{figure}[t]
 \begin{center} 
 \includegraphics[width=0.49\textwidth]{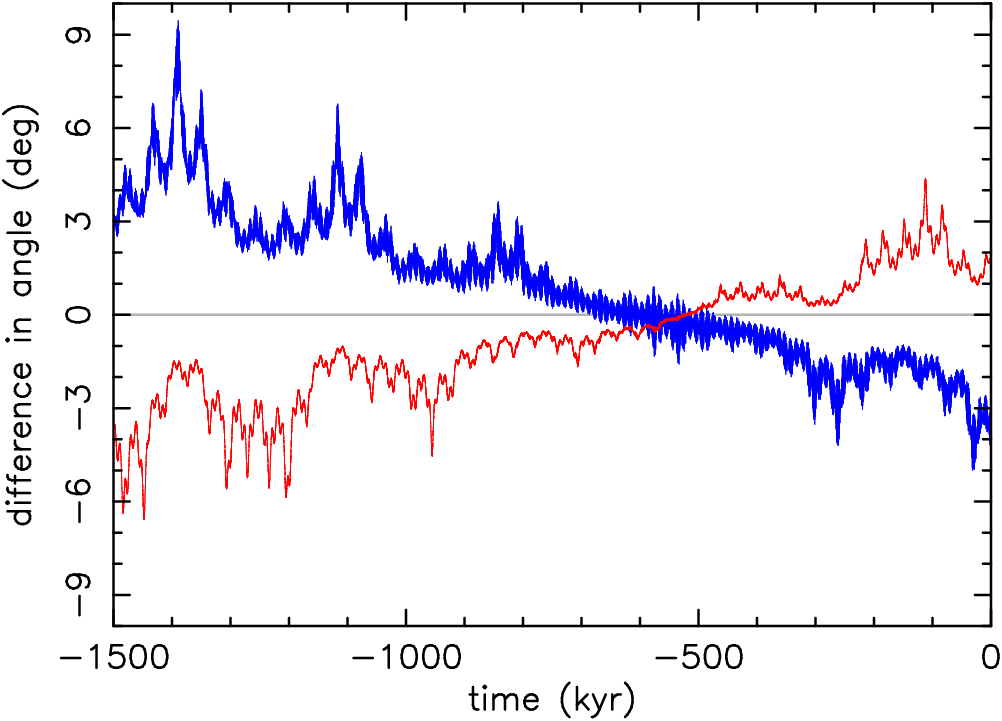}
 \end{center}
 \caption{\label{f2}
  Orbital convergence of secular angles for the nominal orbits of 2016~GO11
  and (525)~Adelaide: (i) difference in longitude of node $\delta
  \Omega=\Omega_{\rm 2016\,GO11}-\Omega_{525}$ (red line), (ii)  difference in longitude
  of perihelion $\delta\varpi=\varpi_{\rm 2016\,GO11}-\varpi_{525}$ (blue line) (osculating values in
  both cases). The abscissa is time to the past (in kyr). At approximately $500$~kyr
  ago, both secular angles of 2016~GO11 converge to the respective values
  of (525)~Adelaide.}
\end{figure}
%%%%%%%%%%%%%%%%%%%%%%%%%%%%%%%%%%%%%%%%%%%%%%%%%%%%%%%%%%%%%%%%%%%%%%%%%%%%%%%%%%%%%%

To further justify membership in the Adelaide family, we numerically propagated
backwards in time the nominal orbits of (525)~Adelaide and the set of $51$ candidate
asteroids from the cluster found in the target box. For the sake of simplicity
we included gravitational attraction by the Sun and perturbations from planets, but
neglected other effects such as the thermal accelerations at this stage. Our
integration spanned a $2$~Myr time interval from the current epoch backwards in time.
We used a well-tested {\tt swift} package%
\footnote{\url{http://www.boulder.swri.edu/~hal/swift.html}}
that efficiently implements N-body propagation using the first order symplectic map
of \citet{wh1991}. We used a short timestep of three days and output the state vectors of
all propagated bodies, planets, and asteroids, every five years for further analysis.
We monitored differences $\delta \Omega=\Omega-\Omega_{525}$ and $\delta\varpi=\varpi
-\varpi_{525}$ between the orbits of identified fragments and (525)~Adelaide. In
contrast to similarly defined differences in semi-major axis, eccentricity, and
inclination, $\delta \Omega$ and $\delta\varpi$ are subject to secular evolution,
clearly detectable on a million-year timescale. This is because proper nodal and pericentre
frequencies $s$ and $g$ are slightly different, even across a compact family such
as Adelaide (see Table~\ref{tab_prop1}). As already mentioned, the present-day values
range in an interval of $\sim \pm 10^\circ$, while the low-velocity dispersal at the
origin of the family must have resulted in $\delta \Omega$ and $\delta\varpi$ not larger
than a fraction of a degree. So, in principle, the current configuration of proposed
asteroids in the Adelaide family is incompatible with the initial state, but
if things work well the initial state might be achieved by past simultaneous convergence
of $\delta \Omega$ and $\delta\varpi$ to near zero values. The benefit of a success in
a similar numerical experiment is twofold;  first, it proves a candidate asteroid is
a true member in the family, and second, the epoch of convergence can be identified
as the origin of the family. This procedure has been used since the discovery
of the Karin family \citep{karin2002}, and applied to very young asteroid
families since the work of \citet{datura2006} and \citet{nv2006}. 
% FIG %%%%%%%%%%%%%%%%%%%%%%%%%%%%%%%%%%%%%%%%%%%%%%%%%%%%%%%%%%%%%%%%%%%%%%%%%%%%%%%%
\begin{figure}[t]
 \begin{center} 
 \includegraphics[width=0.49\textwidth]{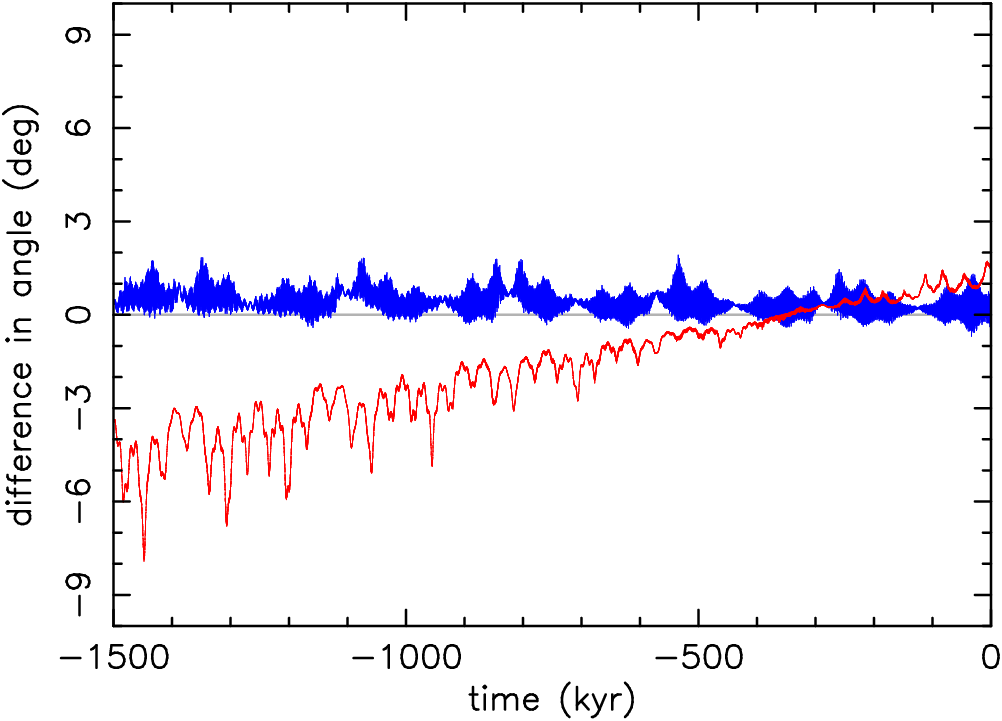}
 \end{center}
 \caption{\label{f3}
  Orbital convergence of secular angles for the nominal orbits of (475474) 2006~SZ152
  and (525)~Adelaide: (i) difference in longitude of node $\delta
  \Omega=\Omega_{475474}-\Omega_{525}$ (red line), (ii) difference in longitude
  of perihelion $\delta\varpi=\varpi_{475474}-\varpi_{525}$ (blue line) (osculating values in
  both cases). The abscissa is time to the past (in kyr). The secular angle differences
  stay very small for the past $\simeq 600$~kyr (within about one degree), but
  the precise nature of the convergence is not as good as in the previous two
  figures  because of the likely contribution of the Yarkovsky acceleration
  in the orbital evolution of the smaller asteroid (475474) 2006~SZ152 (see  text).}
\end{figure}
%%%%%%%%%%%%%%%%%%%%%%%%%%%%%%%%%%%%%%%%%%%%%%%%%%%%%%%%%%%%%%%%%%%%%%%%%%%%%%%%%%%%%%

In order to proceed step-by-step, however, we first perform a less ambitious task.
At this moment we analyse $\delta \Omega$ and $\delta\varpi$ individually for
each of the candidate asteroids in the Adelaide family, without combining them 
into a global picture. Therefore, at this stage we are not aiming to determine
the age of the Adelaide family;  our primary goal is to eliminate interloper objects
from our candidate list. These interlopers will not show any signs of convergence of
$\delta \Omega$ and $\delta\varpi$ to zero at approximately the same time.

An example of a successful convergence test for asteroid (452322) 2000~GG121 is shown in 
Fig.~\ref{f1}. The simultaneous crossing of zero of the angular difference with (525)~Adelaide
occurs at $\simeq 550$~kyr ago, in a close agreement with the suggested age of Adelaide in
\citet{ade2019}. In spite of some oscillatory patterns,%
\footnote{Their principal periods are $\simeq 46$~kyr and $\simeq 300$~kyr, related
 to $g_5$ and $g_6$ frequencies,  because $\delta \Omega$ and $\delta\varpi$ 
 are also subject to the forced terms and, in particular, the $g_6$ term is slightly
 amplified by a $g-g_6$ divisor of the nearby $\nu_6$ secular resonance.} 
both $\delta \Omega$ and $\delta\varpi$ behave approximately linearly in time.
Moreover, the slope of this linear trend closely matches the difference in
proper frequencies $s$ and $g$ determined for (525)~Adelaide and
(452322) 2000~GG121 in the Appendix (Table~\ref{tab_prop1}). For instance, $\delta \Omega\simeq
(s_{452322}-s_{525})\,t$ with $s_{452322}-s_{525}\simeq -4.8\times 10^{-2}\;\arcsec$~yr$^{-1}$
(the positive slope in Fig.~\ref{f1} is explained by time going into the past at the abscissa),
and similarly for $\delta\varpi$. The difference in the proper frequencies is mainly due to the
difference in  the proper semi-major axes of these two asteroids (Table~\ref{tab_prop}), which
is among the largest in the family. It is worth noting that the nodal behaviour 
at the crossing condition $\delta \Omega = 0$ is smooth and well behaved. On the contrary,
even though the $\delta \varpi = 0$ condition occurs at about the same epoch, the
perihelion convergence suffers oscillatory terms. We assume this is due to a slight
offset in proper eccentricity values between (525)~Adelaide and (452322) 2000~GG121
(Table~\ref{tab_prop}). Figure~\ref{f2} shows another successful convergence test, this
time for asteroid 2016~GO11. Its proper semi-major axis is among the smallest in the
family (Table~\ref{tab_prop}),
at the opposite extreme to (452322) 2000~GG121 discussed above. For this reason,
the linear approximation of $\delta \Omega$ and $\delta\varpi$  now have opposite trends.
Nevertheless, both $\delta \Omega \simeq 0$ and $\delta\varpi\simeq 0$ occur again at
about $510$~kyr ago, closely compatible with the proposed age of the family, and again the
perihelion convergence suffers a jitter by periodic terms related to an offset in proper
eccentricity. About one-third of the asteroids in our candidate sample exhibited this
excellent type of secular angle convergence.

Another group of candidate asteroids showed behaviour that was a little worse than
$\delta \Omega$ and $\delta\varpi$, although still acceptable. A representative of this class,
asteroid (475474) 2006~SZ152, is shown in Fig.~\ref{f3}. While  a simultaneous convergence of
nodes and pericenters is not apparent here,  $\delta \Omega$ and $\delta\varpi$ both remain
very small (within a degree limit in this case) for the past $\simeq 600$~kyr. Even the
starting values are currently very small, reflecting the close proximity of the
orbit of (475474) 2006~SZ152 to (525)~Adelaide (see the $a_{\rm P}$ values listed in
Table~\ref{tab_prop}). We are positive that in these cases, convergence near $500$~kyr
ago can be achieved when the effect of thermal accelerations (the Yarkovsky effect)
are included in the simulation. As discussed in Sect.~5 of \citet{datura2017}, the
Yarkovsky effect would present an additional quadratic trend in $\delta \Omega$ and
$\delta\varpi$. For instance, in the nodal case $\delta \Omega_{\rm Yar}\simeq 0.5\,
(\partial s/\partial a)\,{\dot a}\,t^2$, where $(\partial s/\partial a)\simeq 38\;
\arcsec$~yr$^{-1}$~au$^{-1}$ in the Adelaide family zone and ${\dot a}$ is the semi-major
axis secular drift of (475474) 2006~SZ152 due to the Yarkovsky effect. Unfortunately,
${\dot a}$ is not known for any of the Adelaide small members, and given their estimated
size and heliocentric distance, it may be any value within $\pm 6\times 10^{-4}$ au~Myr$^{-1}$
\citep[e.g.][]{betal2006,vetal2015}. With $t\simeq 500$~kyr, the Yarkovsky effect
contribution to the nodal convergence condition may be within the range $\delta
\Omega_{\rm Yar}\simeq \pm 1^\circ$. A similar contribution is also possible to the perihelion
convergence. The Yarkovsky effect may also contribute in the convergence of asteroids
(452322) 2000~GG121 and 2016~GO11 shown in Figs.~\ref{f1} and \ref{f2}, but it clearly
cannot explain the full range of the present-day $\delta \Omega$ and $\delta\varpi$ spanning
several degrees. In these cases simulation without the thermal accelerations have
already provided a nicely consistent picture of the past convergence and it primarily
derives from a distant-enough location of their orbits from (525)~Adelaide.

Finally, (159941) 2005~WV178 was the only case that was found to be orbitally dissimilar to
(525)~Adelaide in the past $2$~Myr. This confirms our earlier suspicion that this object
is unrelated to the Adelaide family, and we further confirm this conclusion by
computation of its proper elements in the Appendix. Our working list of known
members in the Adelaide family is given in Table~\ref{tab_prop}.
% FIG %%%%%%%%%%%%%%%%%%%%%%%%%%%%%%%%%%%%%%%%%%%%%%%%%%%%%%%%%%%%%%%%%%%%%%%%%%%%%%%%
\begin{figure}[t]
 \begin{center} 
 \includegraphics[width=0.49\textwidth]{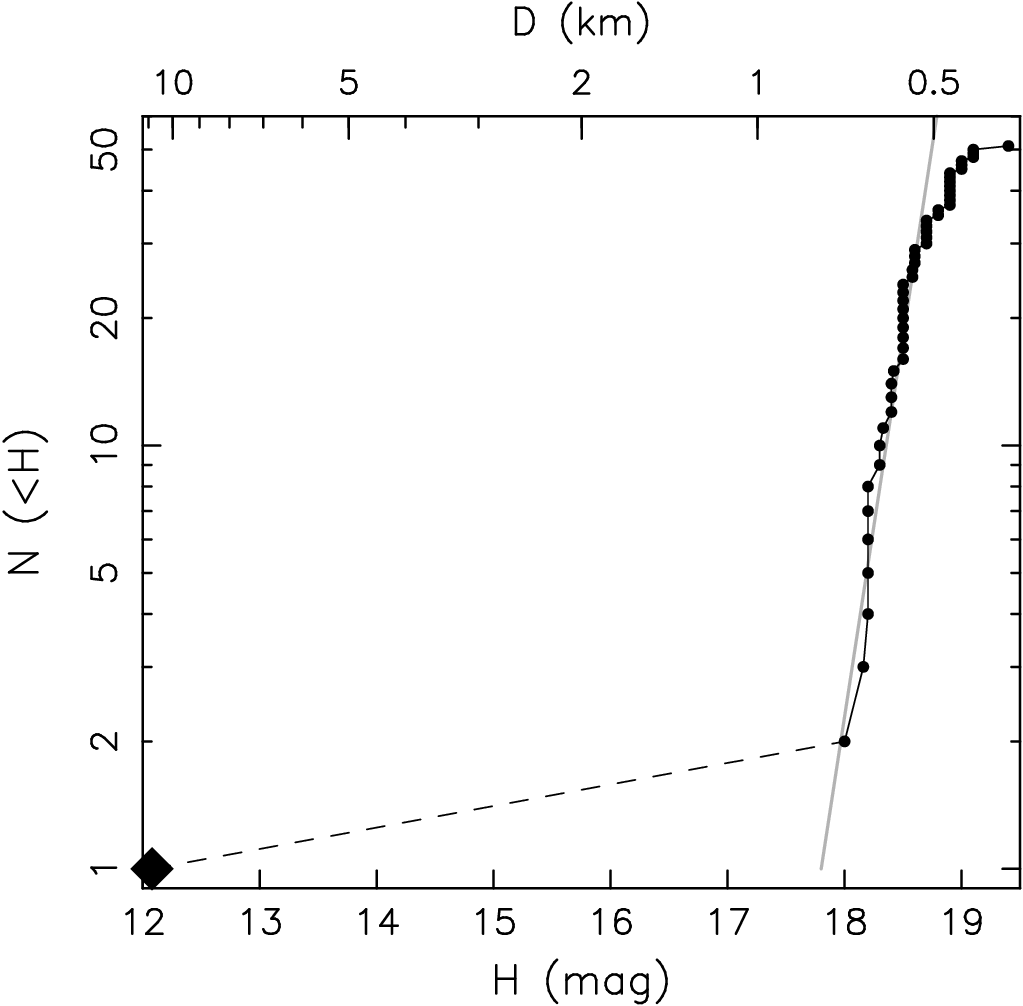}
 \end{center}
 \caption{\label{f4}
  Cumulative distribution $N(<H)$ of absolute magnitude $H$ for known Adelaide family
  members. The diamond symbol denotes the largest fragment (525)~Adelaide, filled
  circles are smaller fragments in the family. The grey line shows the power-law
  relation $N(<H)\propto 10^{\gamma H}$ for $\gamma=1.8$, which approximates the
  distribution of small members. The upper abscissa provides an estimate of
  size, assuming geometric albedo $p_V=0.22$ (a possible value for (525)~Adelaide;
  Sect.~\ref{lf}). The sizes would be slightly smaller for larger $p_V$ values.}
\end{figure}
%%%%%%%%%%%%%%%%%%%%%%%%%%%%%%%%%%%%%%%%%%%%%%%%%%%%%%%%%%%%%%%%%%%%%%%%%%%%%%%%%%%%%%

% SFD
Once we have resolved the question of the Adelaide family membership, we can
analyse several population
characteristics. We start with the simplest case, namely the distribution of absolute
magnitude values (equivalent to the size distribution, if the albedo is constant).
The data in Table~\ref{tab_prop} indicates that (525)~Adelaide is the only sizable
object in the family and all remaining asteroids are much smaller. Assuming $p_V=0.22$
(see Sect.~\ref{lf}), the sizes range between $\simeq 0.43$~km and $\simeq 0.71$~km,
but slightly smaller sizes are also possible if the geometric albedo is actually
larger. Figure~\ref{f4} shows the cumulative distribution $N(<H)$ of absolute magnitude
$H$. Omitting (525)~Adelaide itself, $N(<H)$ can be approximated using a power law $N(<H)
\propto 10^{\gamma H}$ with $\gamma\simeq 1.8$. The extreme size difference between the
largest fragment and
the second largest fragment, and the very steep power-law exponent $\gamma$ imply the
Adelaide family results from a significant cratering event on Adelaide itself. The
$N(<H)$ distribution resembles that of other cratering-born families such as Vesta or
Massalia. It should also be noted  that the slow rotation of (525)~Adelaide
(Sect.~\ref{lf}), and the limited mass of the sub-kilometre-sized fragments, imply
the Adelaide family cannot form by a rotational fission of its parent body
\citep[see][]{petal2018}.

Because the obvious observational incompleteness of sub-kilometre-sized asteroids in
the main belt, we wondered if we were missing some larger Adelaide fragments than those
listed in Table~\ref{tab_prop}. However, we consider this possibility unlikely as
several studies place the observational completeness limit at the Adelaide zone between
$17$ and $18$ magnitude. For instance, \citet{datura2017} evaluated the completeness of
Catalina Sky Survey observations taken  between 2005 and 2012 for Datura family
members (whose  orbits are very similar to those of  Adelaide). They found that the Datura
population is basically
complete at $\simeq 16.8$~mag from just this survey. \cite{hm2020} followed a more empirical
approach,  using all the data combined as of 2020 to estimate population completeness limit
as a simple function of the semi-major axis. They obtained $\simeq 17.5$~mag limit at the
Adelaide zone. As a result, it seems unlikely that we are missing any large fragments
in the Adelaide family. Obviously, the population beyond magnitude $18$ becomes gradually
less complete, and our set of $50$ fragments with $H$ between $18$~mag
and $19$~mag may represent just a fraction of the true population.

\subsection{Proper element determination for the Adelaide cluster} \label{prop}
The family identification in the previous section circumvents proper orbital
elements. However, it is still very useful to determine these elements because
they may reveal some details about family structure. They represent
yet another independent step in the justification for family membership 
because a potentially young family must be extremely compact in the space of
proper orbital elements, even  more than in the respective osculating
elements.

We used synthetic proper elements $(a_{\rm P},e_{\rm P},\sin I_{\rm P})$ defined by
purely numerical means \citep[e.g.][]{km2000,km2003}. A world-wide database
of the asteroid proper elements is maintained at the website {\tt AstDyS},%
\footnote{\url{https://newton.spacedys.com/astdys/}}
run by a consortium of institutions lead by the University of Pisa. However,
the information at {\tt AstDyS} is not up-to-date, and does not provide proper elements
for all the asteroids in the Adelaide family. For this reason we determined the
proper elements ourselves, using the methods described in the Appendix. Results
for $(a_{\rm P},e_{\rm P},\sin I_{\rm P})$ and their formal uncertainty, are provided
in Table~\ref{tab_prop}. Proper frequencies of node and perihelion $(s,g)$ and
their formal uncertainty are listed in Table~\ref{tab_prop1}. In a few cases
of asteroids observed in only two or three oppositions, the formal semi-major
axis uncertainty $\delta a_{\rm P}$ may underestimate the realistic value (see 
the discussion in the Appendix for more examples and details). For
instance, the orbit with the least observational data (2015~TD44, which has only
25 astrometric observations during two oppositions in 2015 and 2020) has an
osculating semi-major axis uncertainty $\delta a\simeq 12.6\times 10^{-6}$~au, about
four times larger than the formal $\delta a_{\rm P}\simeq 3.3\times 10^{-6}$~au
(computed from the nominal osculating orbit). While we admit this slight
inconsistency, we are fortunate that it does not have any impact on our conclusions. We note
that the formal uncertainties on proper $e_{\rm P}$ and $\sin I_{\rm P}$ are always
larger than the uncertainties on the respective osculating $e$ and $I$ for
multi-opposition asteroids. A special category in this respect comprises  the four
single-opposition objects in the family: 2016~UO110, 2017~RS100, 2019~TC62, and
2019~YE29. In these four cases the osculating orbits have uncertainty values that are larger
than the stated values of the respective proper elements; the worst situation
is for 2017~RS100 observed over a  nine-day arc only resulting in an osculating
semi-major axis uncertainty of $\delta a\simeq 7.5\times 10^{-3}$~au. We include
these four objects for the sake of completeness, but again their omission is not
important for any of our conclusions.
% FIG %%%%%%%%%%%%%%%%%%%%%%%%%%%%%%%%%%%%%%%%%%%%%%%%%%%%%%%%%%%%%%%%%%%%%%%%%%%%%%%%
\begin{figure}[t]
 \begin{center} 
 \includegraphics[width=0.49\textwidth]{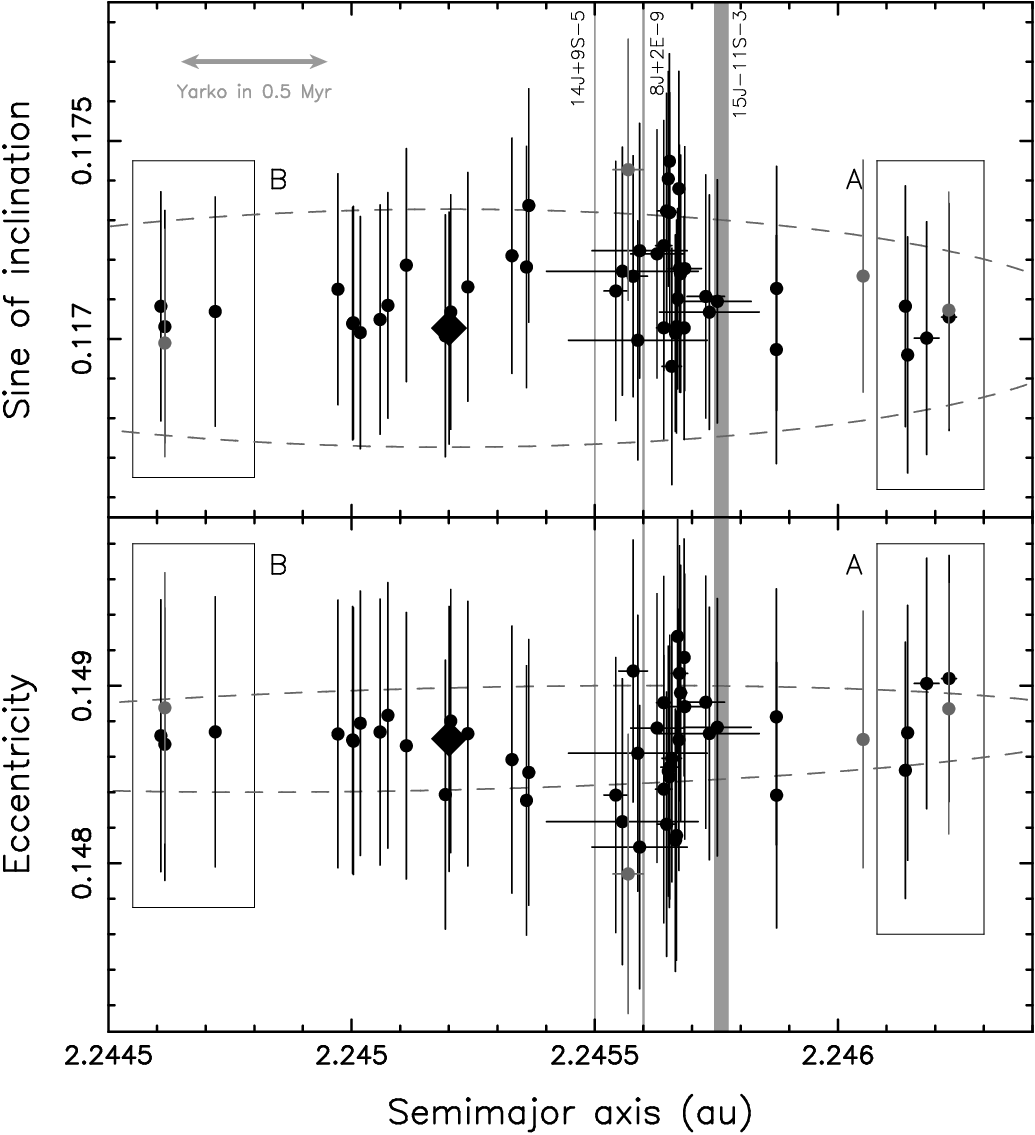}
 \end{center}
 \caption{\label{f5}
  Adelaide family represented by two possible plane-projections of proper orbital
  elements: (i) semi-major axis $a_{\rm P}$ vs sine of inclination $\sin I_{\rm P}$
  (top panel), and (ii) semi-major axis $a_{\rm P}$ vs eccentricity $e_{\rm P}$ (bottom
  panel). Shown are the largest fragment (525)~Adelaide (diamond) and the smaller family
  members (black filled circles  for multi-opposition orbits, grey for single-opposition
  orbits). The vertical and horizontal bars are the formal uncertainty values of the
  proper elements from Table~\ref{tab_prop}. The vertical grey stripes indicate the
  locations of several weak mean motion resonances (from left to right): 14J+9S-5,
  8J+2E-9, and 15J-11S-3; the width of the line approximates resonances strength at
  the value of proper eccentricity using \citet{gallardo2014}. Boxes A and B contain
  asteroids whose orbits are well separated from (525)~Adelaide, suitable for dating
  family origin using backward integrations (see Sect.~\ref{age}). The grey arrow in the
  top panel gives an estimate of maximum drift in semi-major axis of a $\simeq 0.5$~km
  asteroid, typical of many small members in the family (Fig.~\ref{f4}), in $500$~kyr.
  The grey dashed ellipses indicate proper element zones in which fragments may land if
  ejected from (525)~Adelaide with $6$~m~s$^{-1}$ velocity (assuming true anomaly $f=
  90^\circ$ and argument of perihelion such that $\omega + f = 0^\circ$ at origin).}
\end{figure}
%%%%%%%%%%%%%%%%%%%%%%%%%%%%%%%%%%%%%%%%%%%%%%%%%%%%%%%%%%%%%%%%%%%%%%%%%%%%%%%%%%%%%%

Figure~\ref{f5} shows the projection of the Adelaide family members onto a plane defined by two
proper elements, semi-major axis $a_{\rm P}$ vs sine of inclination
$\sin I_{\rm P}$ in the top panel, and  semi-major axis $a_{\rm P}$ vs eccentricity
$e_{\rm P}$ in the bottom panel. As  expected, the family is extremely
compact and all the proposed members reside in the space of proper elements at a
satisfactory distance from the orbit of (525)~Adelaide. This is quantitatively
expressed by a velocity distance of $6$ m~s$^{-1}$ (the lower estimated limit for
the escape velocity of this asteroid; Sect.~\ref{lf}) shown by the dashed ellipses
in both panels of Fig.~\ref{f5}. For the sake of simplicity, the velocity distance
in the proper element space used here assumes an isotropic ejection field, and
requires a choice of the true anomaly $f$ and argument of perihelion $\omega$ at
the epoch of the formation of the family (a priori unknown parameters). We set $f=90^\circ$ and 
$\omega + f = 0^\circ$, which conforms well to the distribution of the Adelaide fragments.
With these characteristics, all the fragments must land inside the dashed elliptical
zones. The fragments at the opposite extreme distances to
(525)~Adelaide in terms of proper semi-major axis are highlighted by boxes A and
B. The asteroid (452322) 2000~GG121, whose nominal orbit convergence is shown in 
Fig.~\ref{f1}, is representative of objects in box A, while the asteroid
2016~GO11, whose nominal orbit convergence is shown in Fig.~\ref{f2}, is representative
of objects in box B. The members in these groups have accumulated the largest
differences in present-day osculating longitude of node and perihelion with
respect to (525)~Adelaide. About 12 additional members lie in the immediate
vicinity of (525)~Adelaide in terms of proper semi-major axis (namely within $\delta
a_{\rm P}\simeq \pm 0.0002$~au). The asteroid (475474)
2006~SZ152, whose nominal orbit convergence is shown in Fig.~\ref{f3}, represents
this class of fragments. Their orbits have small differences in osculating secular
angles from those of (525)~Adelaide. As discussed above, adjusting their exact
convergence critically requires the contribution of the Yarkovsky effect. Finally,
more than half of the small fragments (26 of 50) in the family reside in a
narrow zone delimited by proper semi-major axis values $2.2455$~au and $2.2458$~au.
This region appears anomalous in the Adelaide family because of the statistically
significant concentration of members, and for  two other characteristics:%
\footnote{We exclude the hypothesis of a secondary fragmentation in the Adelaide
 family in this zone  because all fragments here have very similar sizes,
 while a secondary break-up would have produced a population of asteroids with a
 characteristic size distribution observed in the families.}
(i)  in this strip the proper semi-major axis uncertainty has elevated values for
many asteroids (with maximum of $\delta a_{\rm P}\simeq \pm 1.6\times 10^{-4}$~au,
about an order of magnitude higher than in other zones of the family), and (ii) the
proper $e_{\rm P}$ and $\sin I_{\rm P}$ values are the most dispersed. In order
to understand what condition may be special in this zone, we conducted a search in the 
catalogue of mean motion resonances with planets compiled by \citet{gallardo2006} and
\citet{gallardo2014}. None of the prominent resonances crosses the Adelaide family,
but we found a few  very weak three-body resonances \citep[e.g.][]{nm1998a,nm1998b}
located exactly in the suspicious strip of anomalous behaviour. The strongest of them is
15J-11S-3 at $\simeq 2.24576$~au, and there are two even weaker resonances (14J+9S-5 and
8J+2E-9 at $\simeq 2.2455$~au and $\simeq 2.2456$~au; see Fig.~\ref{f5}).
Interaction with these resonances, in particular jumps
between their multiplets, can produce a direct perturbation in proper semi-major axis
and may serve as a pre-requisite of the observed anomaly.

\begin{figure*}[t]
 \begin{center} 
 \includegraphics[width=0.75\textwidth]{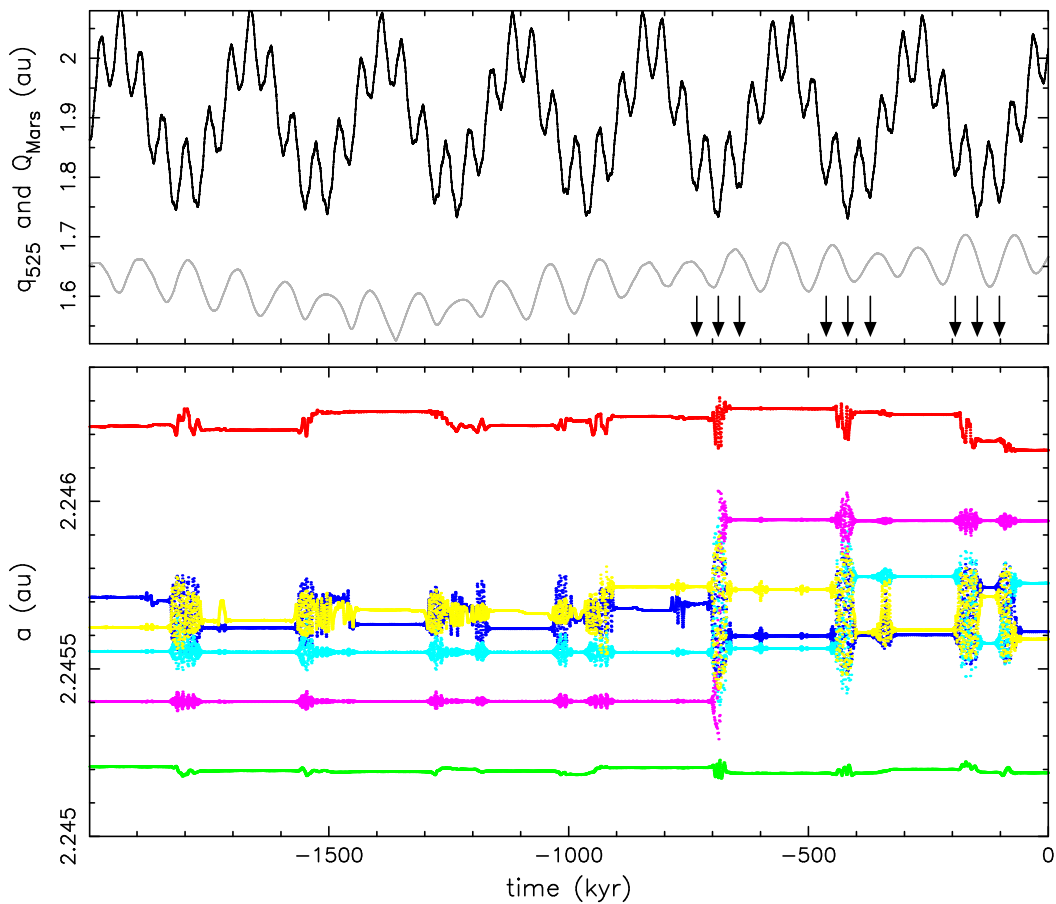}
 \end{center}
 \caption{\label{f6}
  Top panel: Osculating perihelion distance of (525)~Adelaide's nominal orbit (black)
  and the osculating aphelion distance of Mars (grey) vs time into the past. Adelaide's
  orbital behaviour is representative, at this level, of all members in the family. The
  arrows indicate epochs of local minima of Adelaide's perihelion distance if
  $\leq 1.8$~au during the past $700$~kyr. Bottom panel: Mean semi-major axis ${\bar a}$
  of six selected members of the Adelaide family: (452322) 2000~GG121 in red; 
  (475474) 2006~SZ152 in green; 
  (504375) 2007~VV73 in blue; (517580) 2014~UZ170 in cyan; (534611)
  2014~UC204 in magenta; and (545614) 2011~SA45 in yellow. The mean value
  of the semi-major axis was obtained from the osculating value by digitally filtering
  short-period terms (periods $\leq 300$~yr). The distinct jumps in ${\bar a}$ correlate with
  the minima at the perihelion distance (therefore maxima of the orbital eccentricity),
  suggesting that shallow Mars encounters along with resonant perturbations are
  the responsible mechanism.}
\end{figure*}

While inspecting the effect of mean motion resonances, we   noticed yet another
process that operates in the Adelaide family, namely an overall weak chaos due to distant
encounters with Mars. The osculating eccentricity temporarily exceeds $\simeq
0.23$, making the osculating perihelion less than $\simeq 1.73$~au, with a period of
$\simeq 280$~kyr related to the $g-g_6$ frequency (Fig.~\ref{f6}). At the current
epoch, planet Mars is at the peak of its secular eccentricity variations, with its aphelion
periodically exceeding $1.7$~au. The principal period here is $\simeq 95$~kyr, 
corresponding to the planetary frequencies $g_4-g_5$. There are also long-period cycles in the
Mars eccentricity making its maximum perihelion distance reach only $\simeq 1.62$~au
about $1.3$~Myr ago related to the planetary frequencies $g_4-g_3$ (Fig.~\ref{f6}). The
difference between an asteroid's perihelion and the Mars aphelion distance, which
may be currently small
($\leq 0.03$~au), does not necessarily imply close-enough encounters. These
encounters occur in three-dimensional Cartesian space and more well-tuned conditions must be
satisfied. We find of particular importance the value of secular angles that span an interval
up to $10^\circ$, as in the current epoch. The values of secular angles correlate
with the values of the semi-major axis, and this may help explain why the encounter influence
is greater in specific zones of the family. To further illustrate the potential role of Mars
encounters in the Adelaide family, we note that it is located right at the bottom of the
100 Myr scale instability strip shown in Fig.~8 of \citet{mn1999}. The orbits located in this
zone evolve due to the conjoint effect of weak resonances and Mars encounters on
this timescale. The Adelaide family is just barely safe from a more violent evolution.
% FIG %%%%%%%%%%%%%%%%%%%%%%%%%%%%%%%%%%%%%%%%%%%%%%%%%%%%%%%%%%%%%%%%%%%%%%%%%%%%%%%%
%%%%%%%%%%%%%%%%%%%%%%%%%%%%%%%%%%%%%%%%%%%%%%%%%%%%%%%%%%%%%%%%%%%%%%%%%%%%%%%%%%%%%%

In an attempt to understand things in more detail, we selected the orbits of six numbered
members of the Adelaide family and integrated their orbits backwards in time for
$2$~Myr (results shown in Fig.~\ref{f6}). These orbits include some examples from the
critical zone of proper semi-major axes between $2.2455$~au and $2.2458$~au, and also
orbits beyond. In this set of simulations, we also paid particular
attention to the physical distance to Mars, which was monitored every year. We found
that basically all orbits exhibit distant encounters with this planet. The closest
registered encounters were at a distance of $\simeq 0.11$~au , about $18$~Hill radii. All significant
encounters occurred within the past $700$~kyr when Mars had a large-enough perihelion distance
(top panel at Fig.~\ref{f6}). To our surprise, we did not find much closer encounters
for orbits with $a_{\rm P}$ in between $2.2455$~au and $2.2458$~au. For instance, the
orbit of (475474) 2006~SZ152 (in green) has only slightly more distant encounters than
that of (534611) 2014~UC204 (in magenta).  However,  there is a significant difference in the behaviour of their
mean semi-major axes ${\bar a}$:  475474 is markedly stable, with only tiny jitter
in ${\bar a}$;   the orbit of 534611 exhibits the largest jump in ${\bar a}$ at about
$700$~kyr (the epoch of a comparably distant Mars encounter for both orbits). We thus
conclude that the direct effect of the encounter on ${\bar a}$ is small in both cases.
However, what makes the difference is the underlying dynamical landscape. In particular,
even a tiny change in ${\bar a}$ may drive the orbit of 534611 across different
multiplets of the weak mean-motion resonances, depending on the particular phase of
their oscillations. We recall that resonance width is a strong function of orbital
eccentricity \citep[e.g.][]{metal1998,nm1998a}, and this increases their role at the
epochs when perihelion attains minimum values. Thus, the conjoint effect of  the
Mars encounters and the mean-motion resonances is the reason for the macroscopically noticeable
chaos at $2.2455$~au and $2.2458$~au. This also means that this zone acts as an attractor,
capable of capturing the orbits by their chaotic mixing, and this produces their apparent
concentration in the space of proper orbital elements (Fig.~\ref{f5}).

The data in Table~\ref{tab_prop} also indicate that characteristic uncertainties on $e_{\rm P}$
and $\sin I_{\rm P}$ are fairly uniform across the Adelaide family: 
$\simeq 7.5\times 10^{-4}$ and $\simeq 3.0\times 10^{-4}$, respectively. These are not
unusually high values  compared to the global distribution of these uncertainties
in the main belt population \citep[e.g.][]{ketal2002}, but they are so apparent 
in Fig.~\ref{f5} partly because of the extreme compactness of Adelaide family in
these parameters. However, some amplification in $\delta e_{\rm P}$ and $\delta \sin
I_{\rm P}$ is due to the aforementioned effect of distant encounters with Mars.

Finally, we recall that the proper elements are constructed using a conservative model
that includes only the gravitational accelerations of the Sun and planets in an asteroid's
equations of motion. However, sub-kilometre-sized Adelaide fragments might be
affected by thermal accelerations (the Yarkovsky effect) to a noticeable level,
even on the timescale as short as $\simeq 500$~kyr suggested by \citet{ade2019}.
In particular, the Yarkovsky effect may result in a secular drift in proper 
semi-major axis accumulating up to $\simeq \pm 3\times 10^{-4}$~au over
that timescale \citep[e.g.][]{betal2006,vetal2015}. This is shown
with the grey arrows on the top panel of Fig.~\ref{f5}. Therefore, the current
family configuration may not exactly reflect the initial state since some fragments
might have been shuffled left or right in Fig.~\ref{f5}. Over the course of
this migration process some fragments might have also been temporarily trapped in
the zone of mean motion resonances. However, our tests show that this happens only
for slowly drifting fragments. This occurs because none of the resonances crossing the family is
particularly strong and the concentration seen in Fig.~\ref{f5} is truly an apparent
feature of the way   the proper elements are determined.

\subsection{Adelaide family's age} \label{age}
The numerical integrations that served in Sect.~\ref{iden} to support the identification of
the Adelaide family's members, represent a template for the estimation of the family's
age. However, they need to be improved in several respects \citep[see][for further details of the
method]{nv2006}.

Firstly, instead of using the behaviour of secular angles with respect to (525)~Adelaide
for each of the family members individually, we needed to monitor a simultaneous
convergence of orbits for as many members as possible (ideally all of them). We selected
an educated sample of Adelaide members. This not only decreased the
CPU requirements, but also eliminated orbits that would otherwise be problematic.
Our choice is  presented below.

Secondly, the identification of the family's origin from the orbital convergence some tens to
hundreds of thousands of years ago, hinged on how accurately we were able to reconstruct the orbits in
the past. This, in turn, depended on two aspects: (i) the accuracy of the orbits at the
present epoch as reconstructed from the observations, and (ii) the accuracy of the
dynamical model used to propagate the orbits backwards in time. Each of these aspects is
limited and needs to be compensated for in the simulations. The first issue  was
treated by representing the initial conditions for orbit integration using a multitude
of geometrical clones roughly located in the six-dimensional uncertainty ellipsoid
obtained from the orbital determination. The main deficiency of the dynamical model
 is the absence of constraints on the thermal accelerations (the Yarkovsky
effect). Each of the propagated orbits needs to be assigned a value of the
Yarkovsky effect, but these range in the interval of values that can be estimated from
an asteroid's size and heliocentric distance. These are the Yarkovsky clones in our
simulation. In order to simplify our simulations we neglected the geometric clones and 
adopted only the nominal (best-fit) orbits. This is clearly justified by noting that
for small objects, like members of the Adelaide family, the perturbation from the
Yarkovsky effect far surpasses the orbital diversity due to geometrical clones. For
each of the small members in the family, we considered $35$ Yarkovsky clones, each
sampling uniformly semi-major axis drift rate values in the range $\pm 6\times 10^{-4}$
au~Myr$^{-1}$. This value was estimated for $\simeq 0.6$~km  asteroids at
Adelaide's heliocentric distance \citep[e.g.][]{betal2006,vetal2015}. Since the
diurnal variant of the Yarkovsky effect dominates, the negative or positive $da/dt$
values assigned to the Yarkovsky clones imply a retrograde or prograde sense of rotation.
The maximally drifting clones accumulate a change in proper semi-major axis in $500$~kyr,
as shown in the top panel of Fig.~\ref{f5}. For the sake of simplicity, we disregarded the
Yarkovsky clones for (525)~Adelaide  because of its much larger size. Since
the Yarkovsky effect is only represented by a value, namely the semi-major axis
drift-rate, we did not implement its physically rooted, sophisticated representation,
  as was done  in \citet{vmc2000}. We instead represented the Yarkovsky effect using a simple
transverse acceleration, resulting in the required semi-major axis drift, as in 
\cite{nv2006}.

Thirdly, we needed to quantify the success in the convergence of orbits in our numerical experiment.
This was again done using the formula from \cite{nv2006}, who defined a target function
\begin{equation}
 \Delta V = na \sqrt{\left(\sin I \Delta \Omega\right)^2 + 0.5
   \left(e\Delta \varpi\right)^2}\; , \label{tf}
\end{equation}
where $na\simeq 19.9$~km~s$^{-1}$, $e$ and $\sin I$ are orbital eccentricity and inclination
(we considered the osculating values of (525)~Adelaide at a given epoch), and $\Delta \Omega$
and $\Delta \varpi$ are dispersal values of longitude of node and perihelion. These
quantities are defined as $\left(\Delta \Omega\right)^2 = \sum_{ij} \left(\Delta \Omega_{ij}
\right)^2/N$, where $\Delta \Omega_{ij}$ are simple differences in osculating nodal 
longitudes of a particular clone for $i$th and $j$th objects, and $N$ is the number
of pair combinations between the asteroids tested (and similarly for perihelia). The
function $\Delta V$
has a dimension of velocity and approximates a characteristic velocity change between the
orbits at a given time, based only on Gauss equations and the information in secular angles.
At the origin of the Adelaide family, we expect $\Delta V$ at the level (or less) of
the escape velocity from (525)~Adelaide. As a result, we used $\Delta V\leq 6$ m~s$^{-1}$
as a criterion to characterise a successful orbital convergence. 
% FIG %%%%%%%%%%%%%%%%%%%%%%%%%%%%%%%%%%%%%%%%%%%%%%%%%%%%%%%%%%%%%%%%%%%%%%%%%%%%%%%%
\begin{figure}[t]
 \begin{center} 
 \includegraphics[width=0.49\textwidth]{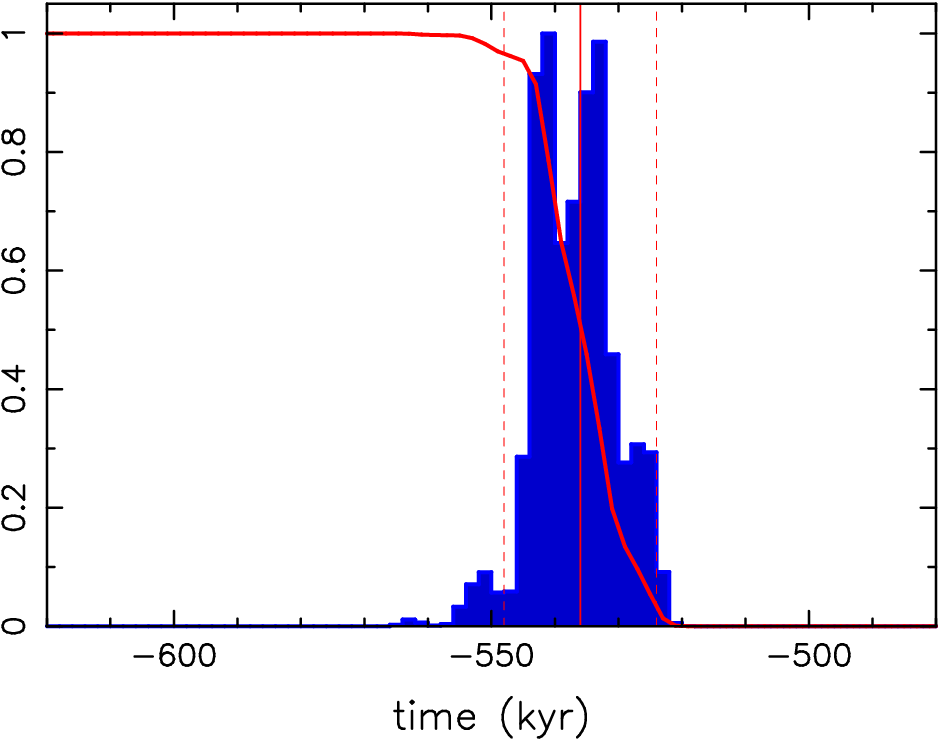}
 \end{center}
 \caption{\label{f7}
  Number of convergent solutions at the $\Delta V' \leq 6$ m~s$^{-1}$ limit from the
  backward integration of six small members with Yarkovsky clones and the nominal orbit
  of (525)~Adelaide. The blue histogram provides differential distribution using
  2 kyr bins (maximum normalised to unity); the red curve is the respective cumulative
  distribution. The thin vertical red line shows the median epoch of the cumulative
  distribution, the vertical dashed lines are 5\%\ and 95\%\ limits. The time is at the
  abscissa to the past.}
\end{figure}
%%%%%%%%%%%%%%%%%%%%%%%%%%%%%%%%%%%%%%%%%%%%%%%%%%%%%%%%%%%%%%%%%%%%%%%%%%%%%%%%%%%%%%

In the course of initial tests we noticed a problem with the perihelion segment in the
target function (\ref{tf}). As already noticed in the examples given in Figs.~\ref{f1} and
\ref{f2}, the longitude in perihelion tends to exhibit short-period oscillations near the 
convergence configuration and their amplitude exceeds the $6$ m~s$^{-1}$ limit on
$\Delta V$. This is due to the sizable scattering of the proper eccentricities of the small
members in the Adelaide family   compared to the proper inclination values. This
behaviour is most likely related to dynamical perturbation due to Mars encounters
that projects in eccentricity more than inclination. Since it is not possible to
exactly reproduce the effects of the Martian encounters in our simulation, we decided to
truncate the contribution from perihelia in the target function (\ref{tf}). We thus used
a simplified version, $\Delta V' = na\,\sin I\, (\Delta \Omega),$ and kept the convergence
criterion $\Delta V'\leq 6$ m~s$^{-1}$. While easier to satisfy, we recall it is still
a very strict limit: plugging in $\sin I \simeq 0.117$ from the respective proper value
(Fig.~\ref{f5}), the $\Delta V' \simeq 6$ m~s$^{-1}$ level corresponds to a nodal dispersion of only
$\Delta \Omega \simeq 0.15^\circ$.
% FIG %%%%%%%%%%%%%%%%%%%%%%%%%%%%%%%%%%%%%%%%%%%%%%%%%%%%%%%%%%%%%%%%%%%%%%%%%%%%%%%%
\begin{figure*}[t]
 \begin{center} 
 \includegraphics[width=0.8\textwidth]{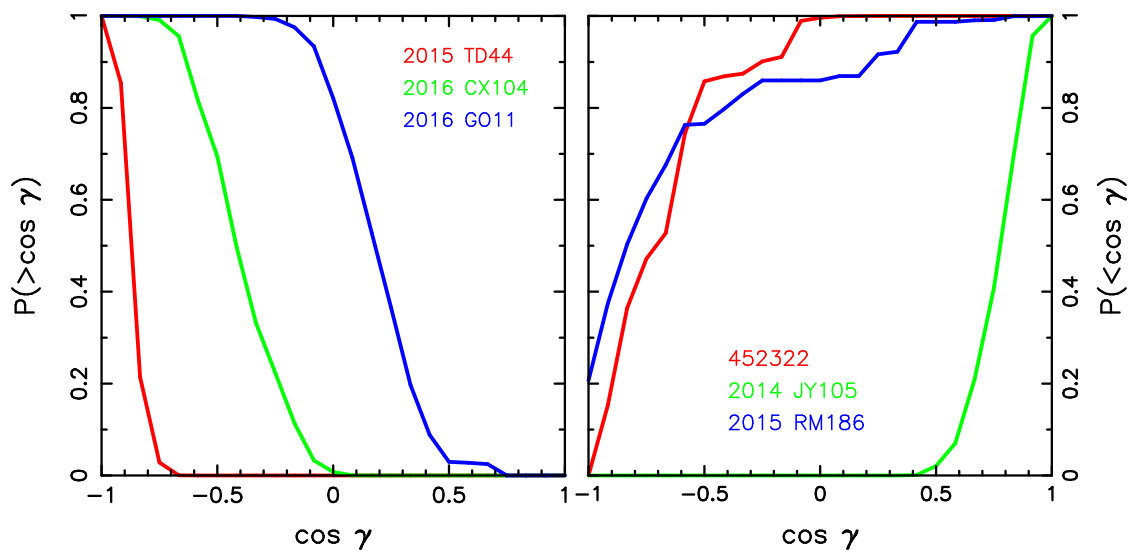}
 \end{center}
 \caption{\label{f8}
  Cumulative distribution of the cosine of obliquity $\cos\gamma$ from the
  converging clones of
  six small Adelaide members in our experiment. The asteroids in   box B (2015~TD44,
  2016~CX104 and 2016~GO11)  are in the left panel; the asteroids in   box A ((452322) 2000~GG121,
  2014~JY105 and 2015~RM186) are in the right panel. The dominantly negative values of
  $\cos\gamma$ (e.g. 2015~TD44 or 2016~CX104) imply a retrograde rotation and migration
  towards the Sun, while the opposite tendency (e.g. 2014~JY105) implies a prograde rotation
  and migration away from the Sun.}
\end{figure*}
%%%%%%%%%%%%%%%%%%%%%%%%%%%%%%%%%%%%%%%%%%%%%%%%%%%%%%%%%%%%%%%%%%%%%%%%%%%%%%%%%%%%%%

We now return to the selection of asteroids used in the numerical simulation targeting at
the age determination for Adelaide family. We included the largest fragment (525)~Adelaide
(nominal orbit and we neglected the 
Yarkovsky effect). In Sect.~\ref{iden}, we identified the objects most distant from
(525)~Adelaide in proper semi-major axis as suitable candidates for family age
determination;  in these cases the dominant part of the secular angle
difference with respect to (525)~Adelaide is simply due to their position in the
family. The badly constrained contribution from the Yarkovsky effect is smaller. In
Fig.~\ref{f5}, we highlight these members with boxes A and B. These are the
objects used in our simulation. In order to alleviate CPU demands, we only excluded
the poorest defined orbit from each of the boxes. As a result the collection of small 
fragments used in our experiment was as follows: (i) in box~A we used (452322) 2000~GG121,
2014~JY105, and 2015~RM186 (excluding 2017~HL72 and 2019~TC62), and (ii) in box~B we used
2015~TD44, 2016~CX104, and 2016~GO11 (excluding 2017~WP50). The other  members in the family
may present difficulties that we intend to avoid in this work. For instance, asteroids too close
to (525)~Adelaide (see Fig.~\ref{f3}) typically have the largest portion of accumulated
difference in secular angles due to the Yarkovsky effect. Even though we consider it in our
simulations, the limited number of Yarkovsky clones may not
guarantee the accuracy of the result. Similarly, all asteroids in the chaotic strip
affected by mean motion resonances and significant perturbations from Mars encounters 
(Sect.~\ref{prop}) may also lead to inaccurate results. 

We were thus left with seven asteroids in our numerical simulation: (525)~Adelaide,
three members   in  box~A (each represented 35 Yarkovsky clones), and three members 
in  box~B (each represented 35 Yarkovsky clones). The simulation covers an interval
of $1$~Myr into the past. Similarly to the simple convergence tests reported in
Sect.~\ref{iden}, we used the {\tt swift} code for orbit propagation. Perturbations from
all planets were taken into account, and here we also implemented thermal accelerations
(the Yarkovsky effect) using a fictitious transverse acceleration providing the
predicted secular change in the semi-major axis \citep[for details of the method,
see][]{nv2006}. The assigned drifts uniformly ranged in the interval $(-(da/dt)_{\rm max},
(da/dt)_{\rm max})$, where $(da/dt)_{\rm max}=6\times 10^{-4}$ au~Myr$^{-1}$ is the
estimated zero-obliquity value for a $0.6$~km asteroid at the Adelaide heliocentric
distance \cite[see][and Sects.~\ref{iden} and \ref{prop} above]{betal2006}. The timestep
was three days and output frequency one year. Every year we consulted the configurations of
all the particles, and for all the possible $35^6\simeq 1.84\times 10^9$ combinations
of the clones of small Adelaide
members and (525)~Adelaide we evaluated the target function $\Delta V'$. When we
encountered a configuration satisfying $\Delta V'\leq 6$ m~s$^{-1}$, we noted the epoch 
and identified the clones that converged. This   information tells us the expected
strength of the Yarkovsky effect, and we interpreted this information in terms of
the predicted sense of rotation.

Figure~\ref{f7} shows the results from our simulation. The median value of the convergence
of the epochs was $536$~kyr. Sorting the convergence epochs in an increasing manner, we obtained
a cumulative distribution of successful clone combinations. We opted to define the age
uncertainty by setting $5$\% and $95$\% limits of the cumulative distribution, which
gave us $536\pm 12$~kyr for the Adelaide family.%
\footnote{The exact quantitative solution for the family's age determination
 depends on what
 level-bar of $\Delta V'$ is used. Our nominal solution uses  $6$ m~s$^{-1}$, motivated by the
 expectation that small fragments are typically launched at the escape velocity from
 the parent body. Luckily, the result does not critically depend on the small variations
 in this criterion. For instance, setting a $\Delta V'=10$ m~s$^{-1}$ limit, allowing thus
 a tail of fast-escaping fragments, would indicate a $538^{+24}_{-16}$~kyr age solution. Much
 higher values of $\Delta V'$ are not expected for cratering events \citep[e.g.][]{setal2017}.}
Additionally, noting which of the clones
contributed to the converging configurations, we  plotted the distributions of the empirical
Yarkovsky drift $da/dt$ assigned to the particular clone for each of the asteroids in
the experiment. Since the diurnal variant of the Yarkovsky effect dominates, we interpreted
this information in terms of the cosine of the obliquity: $\cos\gamma = (da/dt)/(da/dt)_{\rm max}$.
Figure~\ref{f8} shows the cumulative distribution of $\cos\gamma$ for the six small members
in our experiment. Some of them show a clear tendency  towards near extreme positive or negative
$\cos\gamma$ values. For instance, 2015~TD44 and 2016~CX104 from  box~B have
$\cos\gamma$ systematically negative, thus having a predicted retrograde sense of rotation
and inward migration. Similarly, clones of 2014~JY105 from  box~A have a strong preference
for positive $\cos\gamma$ values, thus a prograde rotation and outward migration. These
predictions might be tested and verified observationally, when good enough photometric
observations of these small asteroids are obtained.

% SEC 3 %%%%%%%%%%%%%%%%%%%%%%%%%%%%%%%%%%%%%%%%%%%%%%%%%%%%%%%%%%%%%%%%%%%%%%%%%%%%%
\section{Discussion} \label{disc}
The Adelaide and Datura families are very similar in many respects: (i) they share
nearly identical locations in the space of proper orbital elements, (ii) they have  
overlapping ages, and (iii) their
scales of events are nearly the same (both were massive cratering events on a $10$~km
parent body). Here we examine their origin from two different perspectives: either
two independent events or causally linked events.
\smallskip

\noindent{\it Adelaide and Datura formations from a background population of
 impactors.} Before
we discuss the more speculative case of a possible relation between the Adelaide and Datura
families, we find it important to analyse the probability of their formation by an impact of a
projectile from the background population of asteroids in the main belt. This is 
the standard
view and comparison situation. We used a classical \"Opik-Wetherill approach to estimate the
collision probability of two Keplerian orbits whose secular angles exhibit regular precession
\citep[e.g.][]{o1951,w1967,g1982}. Applying this formulation to the case of a collision between
Datura and Adelaide versus the main belt population, we obtained an intrinsic collisional
probability $P_{\rm i}\simeq 2.9\times 10^{-18}$ km$^{-2}$~yr$^{-1}$ and a mean impact
velocity of $v_{\rm imp}\simeq 5.2$ km~s$^{-1}$ \citep[see also][]{betal2015,betal2020}.
Taking into consideration that both (525)~Adelaide and (1270)~Datura are approximately
$10$~km in size \citep[e.g.][]{datura2009}, the critical impact specific
energy is $Q_{\rm D}^\star\simeq 3000$ J~kg$^{-1}$ \citep[e.g.][]{betal2015,
betal2020}. Together with the $v_{\rm imp}$ stated above, this implies that the impactors
$\geq 580$~m in size (or larger), are capable of catastrophically disrupting Adelaide- or
Datura-size objects \citep[e.g.][]{betal2015}. The smaller impactors produce cratering
events. Since both the Adelaide and 
Datura families represent large cratering events, we assumed an impactor of  $\simeq 100$~m
in size, and noted there are $\simeq 10^8$ such asteroids in the main
belt \citep[e.g.][]{betal2015,
betal2020}. Putting together these data, we found the yearly probability of such an impact
specifically on (525)~Adelaide or (1270)~Datura to be $\simeq 7.2\times 10^{-9}$ yr$^{-1}$
or one such impact every $\simeq 140$~Myr. However, there is nothing special about
(525)~Adelaide or (1270)~Datura, as they are only two of about $\simeq 10^4$ $10$~km-sized
objects in the main belt
\citep[e.g.][]{betal2015}. Any one of them may be the seed of a Datura- or Adelaide-like family.
So, we can presume that these events happen once every
$\simeq 14$~kyr somewhere in the main belt \citep[compare
with Fig.~15 in][]{betal2005}. Since only $\simeq 10-15$~\% of these objects are in the inner
main belt \citep[e.g.][]{metal2011}, the frequency
of the occurrence of the Datura- or Adelaide-like events in the inner main belt, where they can be
more easily discovered, is once per $\simeq 140$~kyr. Given this information, the similar
ages of the two families in that rank is slightly anomalous, but not shocking. What adds a little
more to their anomaly is their orbital proximity, but even that might be a fluke. So, we
 conclude that the formation of the two families independently from background-population
impactors is, in fact, possible.
\smallskip

\noindent{\it Adelaide and Datura formations in causal relation.} We now 
analyse the possibility of a causal relation between the origin of the two families.
We make the assumption that, for instance, the Adelaide family formed first
and the fresh stream of its numerous fragments hit Datura
to form its own family (changing the role of the families in this logic chain does not 
influence the results too much). We again assume that a fragment of $\simeq 100$~m in size
serves as the impactor for the Datura family. The orbital proximity of the two
families implies an HCM distance of only $\simeq 730$ m~s$^{-1}$, implying that a larger Adelaide
fragment would perhaps be needed to produce an   effect similar to that considered above for
background impactors hitting at much higher velocity. This would actually only strengthen
our conclusions. In order to estimate the total number of such fragments in the Adelaide
family we consulted the currently observed population in Fig.~\ref{f4}. Starting from
the second largest fragment of $\simeq 700$~m in size, the cumulative size distribution
is quite steep with a power-law exponent of $\alpha\simeq -(8-9)$. This trend, however, cannot
extend too far because the total mass of this fragment population would soon exceed
that of Adelaide. Hence, at some size the distribution must get shallower. We thus
used a two-slope power-law approximation of the cumulative size distribution of the
Adelaide fragments   between $100$~m and $700$~m, with a very steep gradient at large
sizes down to a breakpoint of $D_{\rm b}$, and a shallower part with a power exponent of $\beta
\geq -3$ below this threshold. Playing with $D_{\rm b}\simeq 400-500$~m and $\beta$, we 
found a number of possible combinations that provide a total mass of $D\geq 100$~m fragments
corresponding to an effective body of $\simeq 2$~km and a cross-section 
corresponding to an effective body of $\leq 10$~km. The total number of $D\geq 100$~m 
fragments remained less than $\simeq (2000-2500)$.
% FIG %%%%%%%%%%%%%%%%%%%%%%%%%%%%%%%%%%%%%%%%%%%%%%%%%%%%%%%%%%%%%%%%%%%%%%%%%%%%%%%%
\begin{figure*}[t]
 \begin{center} 
 \includegraphics[width=0.9\textwidth]{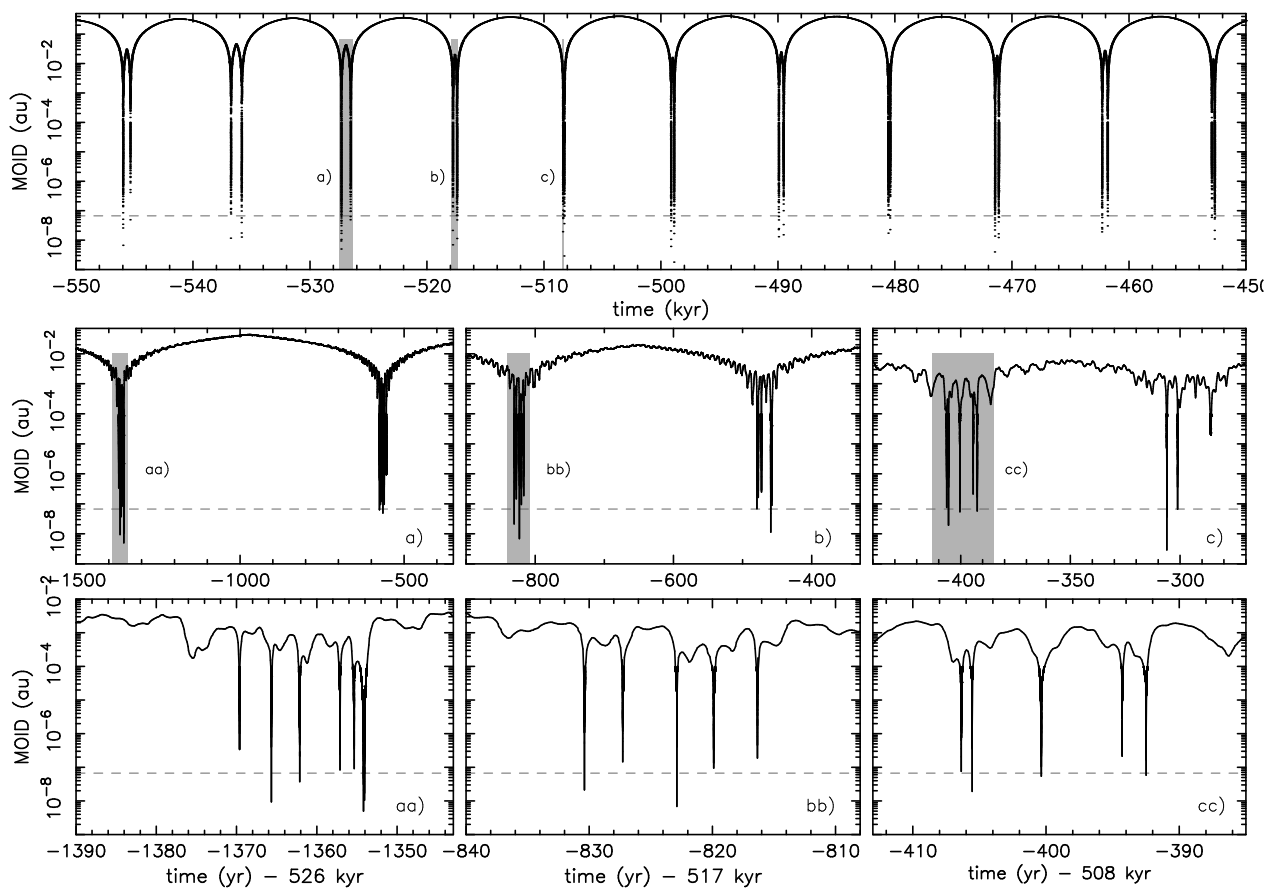}
 \end{center}
 \caption{\label{f9}
  Minimum orbital intersection distance (MOID) for the nominal orbits of (525)~Adelaide and 
 (1270)~Datura, numerically propagated backwards in time (at the  abscissa of all panels).
 The top panel shows the general behaviour of MOID from $450$~kyr to $550$~kyr ago. Zoom-ins on the
 periods shown in the grey rectangles (labelled a  to c) near the minima of MOID are
 shown in the three middle panels, and  provide the information at a greater time resolution. The
 same is repeated in the grey rectangles (labelled aa to cc) in the middle panels, and shown
 at  still greater resolution in the bottom panels. The time in the middle and bottom panels
 is in years and in the top panel in kyr. The horizontal dashed line in all the panels shows
 the MOID $=10$~km threshold.}
\end{figure*}
%%%%%%%%%%%%%%%%%%%%%%%%%%%%%%%%%%%%%%%%%%%%%%%%%%%%%%%%%%%%%%%%%%%%%%%%%%%%%%%%%%%%%%

Because of the extreme proximity of the Datura and Adelaide families, the relative orbital
architecture changes  very slowly. This is, for instance, expressed in the difference
of proper nodal and perihelion frequencies $s$ and $g$, We find that $\delta s\simeq 3.4
\times 10^{-2}\;\arcsec$~yr$^{-1}$ and $\delta g$ is only about an order of magnitude 
greater. The secular difference in proper longitude of node accumulates to $\simeq 5^\circ$
in $500$~kyr and about ten times more in proper longitude of perihelion. 
In spite of the forced contributions in both quantities with an amplitude of
$\simeq 30^\circ-40^\circ$, the difference in the osculating node and perihelion of Datura
and Adelaide accumulate on a  million-year timescale to only a few tens of degrees. The
plain use of the traditional
\"Opik-Wetherill collision-probability estimate is therefore meaningless for our task.
Instead, we use a more straightforward approach. We numerically integrated the nominal orbits
of (525)~Adelaide and (1270)~Datura backwards in time to the  $600$~kyr epoch. The integration
timestep was intentionally chosen to be very short, $0.1$~day. At every timestep, we converted
the state-vectors of both orbits to osculating Keplerian orbits and computed their
minimum orbit intersection distance (MOID). To complete this task efficiently, we used
the algorithm presented in \citet{g2005}, including a Fortran code kindly provided by
the author. If the computed MOID was less than $10^{-4}$~au, we outputted the result
to a file; otherwise, we used a lower output rate of $40$~days. We considered the results from
this numerical test sufficient, at least for our purposes and even for other
possible orbits of fragments related to the Adelaide family.

Figure~\ref{f9} shows our results. The panels from top to bottom provide information 
with an increasing level of resolution, with the middle and bottom panels focusing on
periods of the minima of MOID. All the zoomed panels correspond to epochs immediately following
the nominal formation age of the Adelaide family; however, the nature of the results repeats
during later MOID minima. These events occur with a periodicity of about $9$~kyr, reflecting
the intricate nature of the short- and long-period variations of osculating orbital elements
of both orbits. Following the procedure of a zoom on particular case, we noted 
a complicated nature of MOID oscillations. We estimated how much time during each of
these periods of MOID minimum the two orbits spent at the state of MOID~$\leq 10$~km.
This was the estimated effective cross-section of all Adelaide fragments larger than
$100$~m supposedly capable of forming the Datura family. This number is about 1 day
during these periods, which amounts to little more than $10$~days in $100$~kyr after the
Adelaide family formation. The orbital periods of both (525)~Adelaide and (1270)~Datura are
approximately $1220$~days. Assuming the fragments are perfectly dispersed around Adelaide's
orbit, largely maximising thus the probability of our estimation, we still
need be satisfied
that Datura is located along its orbit right in the $10$-day window interval
(cumulatively). This probability is only about $10/1220\simeq 8.2\times 10^{-3}$.
So, even when overestimating the chances, we found that there is only a little less than a
1\%\  chance that the Datura family formed as a consequence of the causal chain
from Adelaide's fragments
within $\simeq 100$~kyr after the Adelaide family formed. This is in contrast with the
basic certainty that a family such as the Datura family formed within this interval
of time simply because of an impact from the background population of $100$~m-sized
impactors.

% SEC 4 %%%%%%%%%%%%%%%%%%%%%%%%%%%%%%%%%%%%%%%%%%%%%%%%%%%%%%%%%%%%%%%%%%%%%%%%%%%%%
\section{Conclusions} \label{concl}
The Adelaide family belongs to a still rare class of very young asteroid families
(aged $\leq 1$~Myr). It is only the third of this type whose population counts
more than a few  tens of members (together with Datura and Schulhof families). This
makes it an interesting target of analysis. As of February~2021 we have found 51 members
in the Adelaide family sorted into two groups according to  their size: the 
largest fragment  (525)~Adelaide (about $10$~km in size), and the remaining
set of 50 sub-kilometre-sized members (the second largest fragment having an
estimated size of
about $700$~m). The Adelaide family is thus a classical example of a huge cratering
event. The perturbative effect of distant encounters with Mars represents a global
dynamical feature in the location of the Adelaide family. A small strip of
the family is also affected by weak three-body mean-motion resonances. Avoiding
the obviously chaotic regions, we performed a backward orbital integration of selected 
Adelaide members to determine their age by the convergence of their longitude of node. We
found the family to be $536\pm 12$~kyr old.

The multitude of similarities between the Datura and Adelaide families prompted us to
speculate about their related origin. Our analysis, however, tends to reject this
hypothesis. We find the likelihood of this scenario to be very small. In contrast, the
formation of Adelaide- or Datura-scale families by the impact of the background
population of asteroids is very likely on a $\simeq 100$~kyr timescale. So,
it is only by a slight coincidence, though perfectly possible, that their ages are close
to each other. The proximity in the space of proper elements only adds slightly
to their anomaly.
The take-away message from our examination of the Adelaide--Datura relation in their
formation is as follows: It is nearly impossible to beat
the vast background of main belt impactors for small-scale events such as
witnessed by the Adelaide and Datura families. Perhaps only the largest-scale 
family forming events, such as Eos or Themis, could meaningfully influence the
main belt in such a way.

\begin{acknowledgements}
 We are grateful to the anonymous referee, whose numerous suggestions helped to
 improve the original version of this paper. 
 This research was supported by the Czech Science Foundation (grant 18-06083S).
 BN acknowledges the support of the Ministry of Education, Science and Technological
 Development of the Republic of Serbia, contract No.~451-03-68/2020-14/200104.
\end{acknowledgements}

%%%\bibliographystyle{aa}
%%%\bibliography{lit}

\begin{appendix}

\section{Adelaide family:  Membership, proper elements, and proper frequencies}
As  has already been mentioned in the main text,
the proper orbital elements of asteroids are  provided by the {\tt AstDyS}
web service, and more recently also by the Asteroid Families Portal%
\footnote{\url{http://asteroids.matf.bg.ac.rs/fam/}}
\citep[{\tt AFP};][]{retal2017}. Both sets of these elements are computed using
the same synthetic approach developed by \citet{km2000} and \citet{km2003}. The
only difference consists in the dynamical model and integration timespan. While the
elements at the {\tt AFP} are computed in a homogeneous manner (i.e.
using the same dynamical model and integration time-span across the whole asteroid
belt), the elements provided by {\tt AstDyS} are obtained using specific set-ups for
each region of the belt. Despite the availability of these two   sets, there are,
however, two main reasons why we needed to compute new proper elements here: (i) none of the
introduced databases currently provides information about all the potential Adelaide family
members, and (ii) the elements are not obtained using exactly the same procedure for all
asteroids (in the case of {\tt AstDyS}) or the procedure is not well suited for young
families (in the case of {\tt AFP}).

Therefore, our main goal here is to produce a homogeneous set of   proper elements for
all Adelaide family members, from a suitably short integration time span. This  set
of elements is more appropriate to study the structure of young families. Otherwise, the
general outline of the computation of the synthetic proper elements that we applied here
is basically the same as discussed in \citet{km2003}, where  interested readers can find
more details. We only briefly outline the main steps and the particular
choice of the parameters. To build on the expertise of the Pisa group, the proper element
computation was performed using the {\tt ORBIT9} integrator,%
\footnote{\url{http://adams.dm.unipi.it/orbfit/}}
which employs a symplectic single-step method (implicit Runge--Kutta--Gauss) as a starter
and a multi-step predictor for the propagation part of the code \citep[see][]{MilaniNobili1988}.

The first step consists of a numerical propagation of the nominal orbits of all family members
for $2$~Myr using the dynamical model that takes into account the gravitational perturbations of
seven major planets (from Venus to Neptune). The effect of Mercury is  taken into account
indirectly by applying a barycentric correction to the initial conditions. The integration
includes an online digital-filtering procedure, which helps to remove short-period oscillations
(up to about $300$~yr). The code thus provides a time series for the mean orbital elements
for each of the propagated asteroids. In the second step the Fourier analysis methods are
used to remove the forced planetary terms and long-period perturbations from the mean elements,
a procedure that eventually leads to the synthetic proper elements. In order to determine
the proper frequencies of nodal and perihelion longitudes, it is necessary to perform a
linear fit of the corresponding angular variables. To this end, the time series of the mean
angular elements are transformed into real continuous functions simply by adding $2\pi$ times
the number of complete cycles. The output obtained in this way has the same long-term slope as
the original data.

The corresponding formal uncertainties are estimated along with the proper elements using
running box tests. In practice this means segmenting the integrated time interval into
smaller parts, each extending $1$~Myr and shifted by $0.1$~Myr (having thus $11$ such
realisations). Using the method outlined above, we compute proper elements for each of these
sub-intervals. Their standard deviation with respect to the proper element values
computed from the whole $2$~Myr interval provides the formal uncertainty.
While effective and easy to introduce, this approach has its caveats since it does not account
for all the possible sources of error. As a result, the realistic uncertainty of the
proper elements
could be somewhat larger. For instance, we recall that the effect of Mercury is only taken into
account indirectly, by applying a barycentric correction to the initial conditions.
However, based on the results for the ten numbered asteroids in the Adelaide family, we verified
by running a limited set of simulations with Mercury included directly that the effect of
excluding Mercury from the base model is generally less than the formal uncertainties on the
proper elements. We also tested the role of the perturbations of the massive bodies in the
main belt, particularly dwarf planet Ceres and Vesta, and found a slightly greater impact
than that of Mercury. Especially encounters to Vesta, which is located in the same orbital
zone as the Adelaide family, produced
noticeable effects. Luckily, the associated uncertainty on the proper elements remains at the
level of their formal uncertainty. Including Ceres and Vesta into our backward integrations,
testing their convergence of secular angles (see Sect.~\ref{age}), however, exceeds the
computational labour of this work and may need to be checked in the future.

Another source of errors, not taken into account so far, is due to the uncertainty on the
initial data in our numerical integrations (which use just the nominal, best-fit realisation).
While this source of error is typically negligible, in the case of some
poorly determined orbits it could become important. To get an estimate of how important
these uncertainties are, we used 2017~AU38 as a test case. This asteroid is a suitable example
because its osculating orbit has a relatively large uncertainty (this case is nearly identical
to 2015~TD44 mentioned in Sect.~\ref{age}). Based on statistics derived from its orbit
determination, we generated $100$ clones of its initial orbit and computed proper
elements for all of them. The correspondingly derived, standard deviations of the proper elements
from this sample are as follows: $\delta a_{\rm P} \simeq 1.3\times 10^{-5}$~au, $\delta e_{\rm P}
\simeq 9.5\times 10^{-5}$, $\delta \sin I_{\rm P} \simeq 5.9\times 10^{-5}$, $\delta g \simeq 2\times
10^{-3}$ $\arcsec$ yr$^{-1}$, and $\delta s \simeq 3.6\times 10^{-3}$ $\arcsec$ yr$^{-1}$.
Comparing these values with the formal uncertainties of the proper elements for asteroid
2017~AU38 listed in Table~\ref{tab_prop}, we found that the two sets of errors are comparable
only for semi-major axis, while other element errors caused by the uncertainty in the osculating
orbit determination were typically an order of magnitude smaller. Therefore, these sources of
errors could be neglected, except for the extreme cases of the single-opposition orbits of
2016~UO110, 2017~RS100, 2019~TC62, and 2019~YE29. 

Our results are summarised in Tables~\ref{tab_prop} and \ref{tab_prop1}, where data for
the Adelaide family members are given.
We also determined the proper orbital elements of candidate asteroid (159941) 2005~WV178,
excluded from the family as an interloper in Sect.~\ref{iden}. We found $a_{\rm P}=
2.249259\pm 0.000003$~au, $e_{\rm P}=0.14545\pm 0.00107$, and $\sin I_{\rm P}=
0.11716\pm 0.00053$. These values are significantly different from those of the
family members in Table~\ref{tab_prop} and confirm that (159941) 2005~WV178 is
unrelated to the Adelaide family. For instance, its formal HCM velocity distance from
(525)~Adelaide is $\simeq 100$ m~s$^{-1}$, far exceeding the estimated escape
velocity from (525)~Adelaide (Sect.~\ref{lf}).

%%%%%%%%%%%%%%%%%%%%%%%%%%%%%%%%%%%%%%%%%%%%%%%%%%%%%%%%%%%%%%%%%%%%%%%%%%%%%%%%%%%%
\begin{table*}[ht]
\caption{\label{tab_prop}
 Proper orbital elements (and their formal uncertainties) of the Adelaide family members.}
\centering
\begin{tabular}{rlccccccc}
\hline \hline
 \multicolumn{2}{c}{Asteroid} & \rule{0pt}{2ex} $a_{\rm P}$ & $\delta a_{\rm P}$ &
  $e_{\rm P}$ & $\delta e_{\rm P}$ & $\sin I_{\rm P}$ & $\delta \sin I_{\rm P}$ &
  $H$ \\
 & & [au] & [au] & & & & & [mag] \\
\hline
\rule{0pt}{3ex}
   525 & Adelaide   & 2.2452003 & 0.0000072 & 0.14870 & 0.00075 & 0.11703 & 0.00029 & 12.1 \\ 
422494 & 2014~SV342 & 2.2456475 & 0.0000186 & 0.14822 & 0.00074 & 0.11732 & 0.00030 & 18.2 \\ 
452322 & 2000~GG121 & 2.2462281 & 0.0000151 & 0.14904 & 0.00069 & 0.11706 & 0.00029 & 18.4 \\ 
463394 & 2013~GV28  & 2.2450021 & 0.0000032 & 0.14869 & 0.00075 & 0.11704 & 0.00029 & 18.6 \\ 
475474 & 2006~SZ152 & 2.2451926 & 0.0000031 & 0.14839 & 0.00076 & 0.11701 & 0.00031 & 18.3 \\ 
486081 & 2012~UX41  & 2.2455429 & 0.0000241 & 0.14838 & 0.00077 & 0.11712 & 0.00033 & 18.6 \\ 
504375 & 2007~VV73  & 2.2456417 & 0.0000167 & 0.14842 & 0.00075 & 0.11724 & 0.00032 & 18.5 \\ 
517580 & 2014~UZ170 & 2.2456581 & 0.0000198 & 0.14859 & 0.00070 & 0.11693 & 0.00030 & 18.5 \\ 
534611 & 2014~UC204 & 2.2455882 & 0.0001423 & 0.14862 & 0.00078 & 0.11700 & 0.00030 & 18.2 \\ 
545614 & 2011~SA45  & 2.2456659 & 0.0000081 & 0.14813 & 0.00074 & 0.11701 & 0.00025 & 18.3 \\ 
       & 2004~HU76  & 2.2456679 & 0.0000084 & 0.14816 & 0.00070 & 0.11703 & 0.00027 & 18.9 \\ 
       & 2004~HJ85  & 2.2456534 & 0.0000102 & 0.14849 & 0.00074 & 0.11745 & 0.00027 & 19.0 \\
       & 2005~UF193 & 2.2456726 & 0.0000105 & 0.14869 & 0.00074 & 0.11738 & 0.00030 & 18.5 \\
       & 2005~UK370 & 2.2456739 & 0.0000167 & 0.14907 & 0.00072 & 0.11718 & 0.00031 & 18.8 \\
       & 2005~VP83  & 2.2456279 & 0.0000542 & 0.14876 & 0.00076 & 0.11721 & 0.00031 & 18.2 \\
       & 2006~SK449 & 2.2458731 & 0.0000035 & 0.14882 & 0.00072 & 0.11697 & 0.00029 & 18.1 \\ 
       & 2007~VT345 & 2.2452387 & 0.0000008 & 0.14873 & 0.00074 & 0.11713 & 0.00029 & 18.5 \\
       & 2008~ET179 & 2.2457280 & 0.0000385 & 0.14891 & 0.00071 & 0.11711 & 0.00031 & 18.6 \\
       & 2008~US17  & 2.2455787 & 0.0000294 & 0.14908 & 0.00074 & 0.11716 & 0.00030 & 18.3 \\
       & 2008~UR182 & 2.2456539 & 0.0000181 & 0.14854 & 0.00074 & 0.11732 & 0.00030 & 18.2 \\
       & 2009~WJ157 & 2.2458738 & 0.0000013 & 0.14838 & 0.00075 & 0.11713 & 0.00031 & 18.7 \\
       & 2010~UF125 & 2.2456849 & 0.0000342 & 0.14888 & 0.00075 & 0.11718 & 0.00031 & 18.5 \\
       & 2010~VC228 & 2.2456836 & 0.0000070 & 0.14916 & 0.00067 & 0.11703 & 0.00028 & 18.2 \\  
       & 2010~VF260 & 2.2456700 & 0.0000071 & 0.14928 & 0.00066 & 0.11710 & 0.00023 & 18.4 \\
       & 2010~XB115 & 2.2457353 & 0.0001025 & 0.14873 & 0.00071 & 0.11707 & 0.00030 & 18.7 \\
       & 2012~TM342 & 2.2450178 & 0.0000050 & 0.14879 & 0.00074 & 0.11702 & 0.00029 & 19.1 \\
       & 2014~AD31  & 2.2452038 & 0.0000032 & 0.14880 & 0.00074 & 0.11707 & 0.00030 & 18.4 \\
       & 2014~EM164 & 2.2450582 & 0.0000004 & 0.14874 & 0.00075 & 0.11705 & 0.00029 & 18.9 \\
       & 2014~JA2   & 2.2456762 & 0.0000122 & 0.14896 & 0.00072 & 0.11716 & 0.00030 & 18.0 \\
       & 2014~JY105 & 2.2461386 & 0.0000002 & 0.14852 & 0.00072 & 0.11708 & 0.00030 & 18.9 \\
       & 2014~WM167 & 2.2457519 & 0.0000693 & 0.14877 & 0.00073 & 0.11709 & 0.00031 & 18.6 \\
       & 2015~BE285 & 2.2453595 & 0.0000021 & 0.14835 & 0.00076 & 0.11718 & 0.00030 & 18.8 \\
       & 2015~HU72  & 2.2455920 & 0.0000978 & 0.14809 & 0.00080 & 0.11722 & 0.00032 & 18.6 \\
       & 2015~RM186 & 2.2461823 & 0.0000252 & 0.14901 & 0.00071 & 0.11700 & 0.00029 & 18.5 \\
       & 2015~TD44  & 2.2446159 & 0.0000033 & 0.14867 & 0.00077 & 0.11703 & 0.00029 & 19.0 \\ 
       & 2015~UR18  & 2.2456509 & 0.0000109 & 0.14852 & 0.00070 & 0.11740 & 0.00027 & 18.9 \\
       & 2015~XZ90  & 2.2450035 & 0.0000054 & 0.14869 & 0.00075 & 0.11704 & 0.00029 & 18.5 \\
       & 2016~AL322 & 2.2453643 & 0.0000108 & 0.14851 & 0.00075 & 0.11734 & 0.00029 & 18.7 \\
       & 2016~CX104 & 2.2446077 & 0.0000017 & 0.14872 & 0.00077 & 0.11708 & 0.00029 & 18.9 \\
       & 2016~FA34  & 2.2453296 & 0.0000112 & 0.14858 & 0.00075 & 0.11721 & 0.00030 & 18.6 \\
       & 2016~GO11  & 2.2447194 & 0.0000018 & 0.14874 & 0.00076 & 0.11707 & 0.00029 & 18.5 \\
       & 2016~QE71  & 2.2455566 & 0.0001558 & 0.14823 & 0.00080 & 0.11717 & 0.00031 & 18.4 \\
       & 2017~AU38  & 2.2456416 & 0.0000150 & 0.14890 & 0.00071 & 0.11703 & 0.00028 & 18.5 \\
       & 2017~HL72  & 2.2461432 & 0.0000007 & 0.14873 & 0.00072 & 0.11696 & 0.00030 & 18.9 \\
       & 2017~TG26  & 2.2450745 & 0.0000020 & 0.14883 & 0.00075 & 0.11708 & 0.00028 & 18.7 \\
       & 2017~UF65  & 2.2451124 & 0.0000014 & 0.14866 & 0.00075 & 0.11719 & 0.00029 & 19.1 \\
       & 2017~WP50  & 2.2449717 & 0.0000034 & 0.14873 & 0.00075 & 0.11713 & 0.00029 & 19.0 \\ [2pt]
   & \e{2016~UO110} & \e{2.2455685} & \e{0.0000310} & \e{0.14794} & \e{0.00079} & \e{0.11743} & \e{0.00033} & \e{18.9} \\
   & \e{2017~RS100} & \e{2.2446158} & \e{0.0000025} & \e{0.14888} & \e{0.00076} & \e{0.11699} & \e{0.00029} & \e{19.1} \\
   & \e{2019~TC62}  & \e{2.2462278} & \e{0.0000033} & \e{0.14887} & \e{0.00070} & \e{0.11707} & \e{0.00030} & \e{18.9} \\
   & \e{2019~YE29}  & \e{2.2460516} & \e{0.0000007} & \e{0.14870} & \e{0.00072} & \e{0.11716} & \e{0.00029} & \e{19.4} \\ [2pt]
\hline
\end{tabular}
\tablefoot{The Adelaide family membership as of February 2021. The first column lists the
 asteroid number (if numbered) and identification. The next six columns provide the asteroid's
 proper elements $(a_{\rm P},e_{\rm P},\sin I_{\rm P})$ and their formal
 uncertainties $(\delta a_{\rm P},\delta e_{\rm P},\delta \sin I_{\rm P})$ determined
 by the methods described in the Appendix. The last column gives the absolute magnitude
 $H$ from the MPC database. Being a by-product of an orbit determination procedure
 from observations of sky surveys, the listed $H$ values might be uncertain.
 The asteroids whose data are listed in roman font are multi-opposition, while the last
 four (in italics) are single-opposition. In the latter cases, the uncertainties
 on the proper elements are only formal since the uncertainty of the osculating
 elements may currently be  higher.}
\end{table*}
%%%%%%%%%%%%%%%%%%%%%%%%%%%%%%%%%%%%%%%%%%%%%%%%%%%%%%%%%%%%%%%%%%%%%%%%%%%%%%%%%%%%

%%%%%%%%%%%%%%%%%%%%%%%%%%%%%%%%%%%%%%%%%%%%%%%%%%%%%%%%%%%%%%%%%%%%%%%%%%%%%%%%%%%%
\begin{table*}[ht]
\caption{\label{tab_prop1}
 Proper frequencies of longitude of node and perihelion (and their formal uncertainties)
 of the Adelaide family members.}
\centering
\begin{tabular}{rlcccc}
\hline \hline
 \multicolumn{2}{c}{Asteroid} & \rule{0pt}{2ex} $g$ & $\delta g$ & $s$ & $\delta s$ \\
 & & [$\arcsec$ yr$^{-1}$] & [$\arcsec$ yr$^{-1}$] & [$\arcsec$ yr$^{-1}$] & [$\arcsec$ yr$^{-1}$]  \\
\hline
\rule{0pt}{3ex}
   525 & Adelaide   & 32.912 & 0.019 & -36.938 & 0.017 \\ 
422494 & 2014~SV342 & 32.910 & 0.016 & -36.934 & 0.017 \\ 
452322 & 2000~GG121 & 32.945 & 0.017 & -36.986 & 0.017 \\ 
463394 & 2013~GV28  & 32.905 & 0.018 & -36.932 & 0.016 \\ 
475474 & 2006~SZ152 & 32.912 & 0.019 & -36.926 & 0.017 \\ 
486081 & 2012~UX41  & 32.919 & 0.019 & -36.935 & 0.018 \\ 
504375 & 2007~VV73  & 32.915 & 0.013 & -36.941 & 0.017 \\ 
517580 & 2014~UZ170 & 32.925 & 0.018 & -36.952 & 0.017 \\ 
534611 & 2014~UC204 & 32.926 & 0.023 & -36.945 & 0.023 \\ 
545614 & 2011~SA45  & 32.914 & 0.012 & -36.935 & 0.019 \\ 
       & 2004~HU76  & 32.917 & 0.014 & -36.936 & 0.018 \\ 
       & 2004~HJ85  & 32.903 & 0.014 & -36.945 & 0.017 \\
       & 2005~UF193 & 32.911 & 0.013 & -36.953 & 0.016 \\
       & 2005~UK370 & 32.919 & 0.014 & -36.969 & 0.017 \\
       & 2005~VP83  & 32.918 & 0.017 & -36.953 & 0.019 \\
       & 2006~SK449 & 32.937 & 0.019 & -36.966 & 0.017 \\ 
       & 2007~VT345 & 32.910 & 0.018 & -36.940 & 0.016 \\
       & 2008~ET179 & 32.926 & 0.015 & -36.965 & 0.016 \\
       & 2008~US17  & 32.919 & 0.017 & -36.965 & 0.018 \\
       & 2008~UR182 & 32.912 & 0.014 & -36.946 & 0.018 \\
       & 2009~WJ157 & 32.932 & 0.019 & -36.947 & 0.017 \\
       & 2010~UF125 & 32.921 & 0.017 & -36.960 & 0.018 \\
       & 2010~VC228 & 32.918 & 0.013 & -36.977 & 0.016 \\  
       & 2010~VF260 & 32.911 & 0.010 & -36.982 & 0.019 \\
       & 2010~XB115 & 32.930 & 0.016 & -36.958 & 0.015 \\
       & 2012~TM342 & 32.906 & 0.018 & -36.936 & 0.016 \\
       & 2014~AD31  & 32.911 & 0.018 & -36.942 & 0.016 \\
       & 2014~EM164 & 32.906 & 0.018 & -36.935 & 0.016 \\
       & 2014~JA2   & 32.919 & 0.014 & -36.964 & 0.016 \\
       & 2014~JY105 & 32.941 & 0.019 & -36.962 & 0.017 \\
       & 2014~WM167 & 32.929 & 0.016 & -36.959 & 0.015 \\
       & 2015~BE285 & 32.913 & 0.019 & -36.928 & 0.017 \\
       & 2015~HU72  & 32.918 & 0.022 & -36.924 & 0.022 \\
       & 2015~RM186 & 32.945 & 0.019 & -36.983 & 0.018 \\
       & 2015~TD44  & 32.892 & 0.018 & -36.918 & 0.016 \\ 
       & 2015~UR18  & 32.906 & 0.015 & -36.947 & 0.017 \\
       & 2015~XZ90  & 32.905 & 0.018 & -36.931 & 0.016 \\
       & 2016~AL322 & 32.909 & 0.019 & -36.934 & 0.016 \\
       & 2016~CX104 & 32.891 & 0.018 & -36.919 & 0.016 \\
       & 2016~FA34  & 32.911 & 0.019 & -36.936 & 0.017 \\
       & 2016~GO11  & 32.895 & 0.018 & -36.924 & 0.016 \\
       & 2016~QE71  & 32.919 & 0.024 & -36.927 & 0.024 \\
       & 2017~AU38  & 32.920 & 0.013 & -36.963 & 0.016 \\
       & 2017~HL72  & 32.945 & 0.019 & -36.971 & 0.018 \\
       & 2017~TG26  & 32.906 & 0.018 & -36.939 & 0.016 \\
       & 2017~UF65  & 32.904 & 0.018 & -36.933 & 0.016 \\
       & 2017~WP50  & 32.901 & 0.018 & -36.931 & 0.016 \\ [2pt]
   & \e{2016~UO110} & \e{32.910} & \e{0.018} & \e{-36.917} & \e{0.019} \\
   & \e{2017~RS100} & \e{32.894} & \e{0.018} & \e{-36.926} & \e{0.016} \\
   & \e{2019~TC62}  & \e{32.944} & \e{0.018} & \e{-36.979} & \e{0.018} \\
   & \e{2019~YE29}  & \e{32.936} & \e{0.019} & \e{-36.965} & \e{0.017} \\ [2pt]
 \hline
\end{tabular}
\tablefoot{The third and fourth columns give the proper perihelion frequency $g$ and its
 formal uncertainty; the fifth and sixth columns give the proper nodal frequency $s$
 and its formal uncertainty. The asteroids whose data are listed in roman font are
 multi-opposition, while the last four (in italics) are single-opposition.}
\end{table*}
%%%%%%%%%%%%%%%%%%%%%%%%%%%%%%%%%%%%%%%%%%%%%%%%%%%%%%%%%%%%%%%%%%%%%%%%%%%%%%%%%%%%

\end{appendix}

\end{document}